\documentclass[sigconf]{acmart}
\usepackage{enumitem}
\usepackage{multirow}

\microtypesetup{protrusion=true, expansion=true}
\AtBeginDocument{%
  }

\copyrightyear{2026}
\acmYear{2026}
\setcopyright{cc}
\setcctype{by}
\acmConference[CHI '26]{Proceedings of the 2026 CHI Conference on Human Factors in Computing Systems}{April 13--17, 2026}{Barcelona, Spain}
\acmBooktitle{Proceedings of the 2026 CHI Conference on Human Factors in Computing Systems (CHI '26), April 13--17, 2026, Barcelona, Spain}
\acmDOI{10.1145/3772318.3790701}
\acmISBN{979-8-4007-2278-3/2026/04}

\begin{document}

%%
%% The "title" command has an optional parameter,
%% allowing the author to define a "short title" to be used in page headers.
\title{Adaptive Bounded-Rationality Modeling of Early-Stage Takeover in Shared-Control Driving}
%%
%% The "author" command and its associated commands are used to define
%% the authors and their affiliations.
%% Of note is the shared affiliation of the first two authors, and the
%% "authornote" and "authornotemark" commands
%% used to denote shared contribution to the research.
\author{Jian Sun}
\email{sunjian@tongji.edu.cn}
\orcid{1234-5678-9012}
\affiliation{%
  \institution{Key Laboratory of Road and Traffic Engineering, Ministry of Education}
  \institution{Tongji University}
  \city{Shanghai}
  \country{China}
}

\author{Xiyan Jiang}
\email{jiangxiyan@tongji.edu.cn}
\orcid{1234-5678-9012}
\affiliation{%
  \institution{Key Laboratory of Road and Traffic Engineering, Ministry of Education}
  \institution{Tongji University}
  \city{Shanghai}
  \country{China}
}

\author{Xiaocong Zhao}
\email{zhaoxc@tongji.edu.cn}
\authornote{Corresponding author.}
\affiliation{%
  \institution{Key Laboratory of Road and Traffic Engineering, Ministry of Education}
  \institution{Tongji University}
  \city{Shanghai}
  \country{China}
}

\author{Jie Wang}
\email{wjie@tongji.edu.cn}
\affiliation{%
  \institution{Key Laboratory of Road and Traffic Engineering, Ministry of Education}
  \institution{Tongji University}
  \city{Shanghai}
  \country{China}
}

\author{Peng Hang}
\email{hangpeng@tongji.edu.cn}
\affiliation{%
  \institution{Key Laboratory of Road and Traffic Engineering, Ministry of Education}
  \institution{Tongji University}
  \city{Shanghai}
  \country{China}}
  
\author{Zirui Li}
\email{zirui.li@ntu.edu.sg}
\affiliation{%
  \institution{School of Mechanical and Aerospace Engineering}
  \institution{Nanyang Technological University}
  \city{Singapore}
  \country{Singapore}}

%%
%% By default, the full list of authors will be used in the page
%% headers. Often, this list is too long, and will overlap
%% other information printed in the page headers. This command allows
%% the author to define a more concise list
%% of authors' names for this purpose.
% \renewcommand{\shortauthors}{Jian Sun, Xiyan Jiang, Xiaocong Zhao, Jie Wang, Peng Hang, and Zirui Li}

%%
%% The abstract is a short summary of the work to be presented in the
%% article.
\begin{abstract}
  Human drivers’ control quality in the first seconds after a handover is critical to shared-driving safety; potentially unsafe steering or pedal inputs therefore require detection and correction by the automated vehicle’s safety-fallback system. Yet performance in this window is vulnerable because cognitive states fluctuate rapidly, causing purely rationality-driven, cognition-unaware models to miss early control dynamics. We present an interpretable driver model grounded in bounded rationality with online adaptation that predicts early-stage control quality. We encode boundedness by embedding cognitive constraints in reinforcement learning and adapt latent cognitive parameters in real time via particle filtering from observations of driver actions. In a vehicle-in-the-loop study (n=41), we evaluated 
  % the model on both 
  predictive performance and physiological validity. The adaptive model not only anticipated hazardous takeovers with higher coverage and longer lead times than non-adaptive baselines but also demonstrated strong alignment between inferred cognitive parameters and real-time eye-tracking metrics. 
  % Specifically, estimated perceptual noise tracked gaze entropy and fixation instability, while inferred looming aversion aligned with saccadic bursts and pupil dilation. 
  These results confirm that the model captures genuine fluctuations in driver risk perception, enabling timely and cognitively grounded assistance.
\end{abstract}

%%
%% The code below is generated by the tool at http://dl.acm.org/ccs.cfm.
%% Please copy and paste the code instead of the example below.
%%
\begin{CCSXML}
<ccs2012>
   <concept>
       <concept_id>10003752.10010070.10010071.10010261</concept_id>
       <concept_desc>Theory of computation~Reinforcement learning</concept_desc>
       <concept_significance>500</concept_significance>
       </concept>
   <concept>
       <concept_id>10003120.10003121.10003122.10003332</concept_id>
       <concept_desc>Human-centered computing~User models</concept_desc>
       <concept_significance>500</concept_significance>
       </concept>
 </ccs2012>
\end{CCSXML}

\ccsdesc[500]{Theory of computation~Reinforcement learning}
\ccsdesc[500]{Human-centered computing~User models}

% % --- CCS Concepts typesetting: keep the leading bullet attached to the label ---
% % This avoids line breaks that leave a lone bullet at the end of a line.
% \makeatletter
% \expandafter\gdef\csname CCS@General@Theory of computation\endcsname{%
%   \mbox{\textbullet~\textbf{Theory of computation}}}
% \expandafter\gdef\csname CCS@General@Human-centered computing\endcsname{%
%   \mbox{\textbullet~\textbf{Human-centered computing}}}
% \makeatother

%%
%% Keywords. The author(s) should pick words that accurately describe
%% the work being presented. Separate the keywords with commas.
% \keywords{Do, Not, Use, This, Code, Put, the, Correct, Terms, for,
%   Your, Paper}
\keywords{Bounded Rationality Modeling, Takeover in Shared-Control Driving, Autonomous Vehicle, Driving Safety}
%% A "teaser" image appears between the author and affiliation
%% information and the body of the document, and typically spans the
%% page.
\begin{teaserfigure}
  \includegraphics[width=\textwidth]{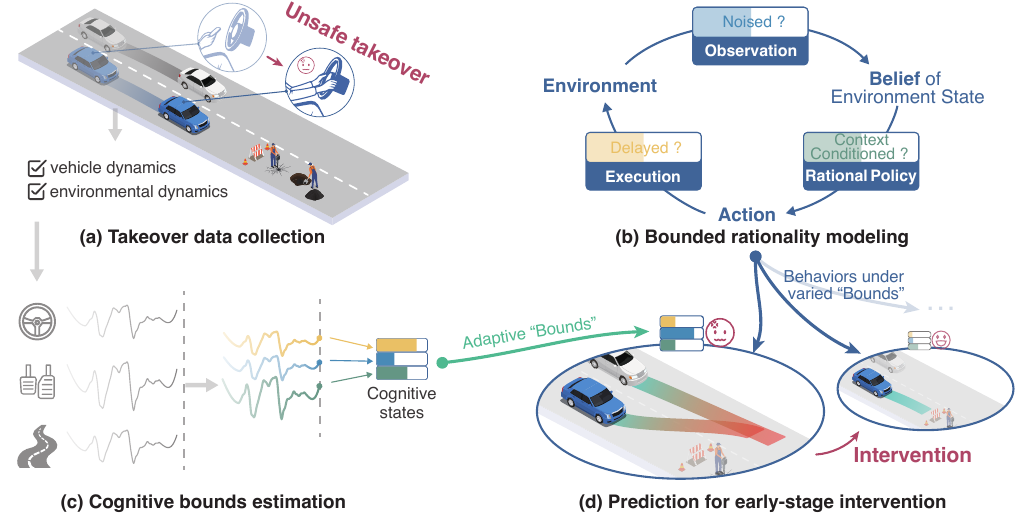}
  \caption{Research framework. (a) Collect multimodal data centered on the takeover moment for model development; (b) Model bounded rationality to describe how drivers act under varied cognitive limitations; (c) Adapt the model with driver’s real-time cognitive state estimation; (d) The adaptive model predicts possible futures, flagging unsafe takeovers to support timely assistance.}
  \Description{A four-part research framework diagram for takeover driving. Panel (a) shows multimodal data collection centered on the takeover moment. Panel (b) depicts a bounded-rationality driver model with cognitive limitations. Panel (c) shows real-time estimation of the driver’s cognitive state to adapt the model online. Panel (d) illustrates future-behavior prediction and the system flagging unsafe takeovers for timely assistance.}
  \label{fig:teaser}
\end{teaserfigure}

% \received{20 February 2007}
% \received[revised]{12 March 2009}
% \received[accepted]{5 June 2009}

%%
%% This command processes the author and affiliation and title
%% information and builds the first part of the formatted document.
\maketitle

\section{Introduction}
Semi-automated vehicles continue to experience preventable safety incidents during control handover. When automation fails or encounters an emergency, drivers may not react quickly or skillfully enough to avoid hazards. Public incident reports show thousands of crashes involving Level-2 driving systems, where automation is active but human drivers remain responsible, in recent years, underscoring the residual risk during takeover~\cite{NHTSA_SGO_2025}. Hands-on-wheel and eyes-on-road provisions are standard, yet prior evidence indicates that these proxies are weak correlates of a safe takeover~\cite{Victor2018AEM, Pipkorn2021HoW}. Human–computer interaction has a stake in this problem because handovers are interaction-critical moments: assistance should align with the person’s tempo and limitations rather than only with system timing.

Predicting the quality of early-stage control—the first seconds after a handover—is therefore essential. In this interval, authority shifts abruptly from automation to the driver, who must manage vehicle dynamics under time pressure and uncertainty. Anticipating imminent control tendencies enables assistance that is timely, well-calibrated, and intelligible, reducing after-the-fact corrections and unnecessary overrides while catching genuinely hazardous cases.

This prediction problem is challenging because cognitive states fluctuate rapidly at handover onset and vary across drivers. Mainstream behavior-prediction methods face two limitations that hinder short-horizon prediction exactly when the system needs it most. First, deep models depend on large labeled datasets and struggle when post-handover data are scarce. Second, they often assume stable, near-rational control, overlooking non-stationarity under time pressure and substantial inter-individual differences. In contrast, the principle of bounded rationality acknowledges that drivers make decisions under strict constraints of information, time, and cognitive capacity. These limitations in human processing amplify performance variability precisely at handover onset.

Recent studies relax strict rationality by incorporating cognitive mechanisms and treating drivers as boundedly rational decision-makers. Prior traffic-interaction research~\cite{ markkula2023explaining, wang2025pedestrian, wang2025modeling, tian2022explaining} and user-simulation studies~\cite{moon2022speeding} in human–computer interaction show that embedding cognitive constraints can improve both individualization and interpretability of modeled behavior. However, many such approaches specify cognition at a psychological or perceptual level without an operational link to executable steering and pedal sequences in closed loop. As a result, cognitive shortfalls are not mapped to concrete safety outcomes, and static parameterizations underfit the fast transients that characterize driver takeover.

We address this gap with a cognition-to-control coupling framework that links specific cognitive limits to executable control and adapts them online during takeover. 
The model instantiates three documented mechanisms: perceptual uncertainty~\cite{faisal2008noise, kwon2015unifying}, referring to noise in estimating surrounding traffic states; looming-averse risk appraisal~\cite{delucia2008critical,tian2022explaining}, reflecting heightened threat sensitivity to rapidly approaching obstacles; and reaction delay~\cite{kosinski2008literature, kuang2015does}, capturing latency between intention and executed steering or pedal actions. These mechanisms are embedded within a partially observable decision process so that cognition modulates what is perceived, how responses to risk are shaped, and when commands are executed. Interpretability in our framework is structural rather than post-hoc: each latent parameter corresponds to a specific cognitive mechanism and causally shapes steering and pedal commands. Building on this structure, a particle-filter layer adapts the latent cognitive parameters in real time from observed actions, enabling personalized, short-horizon prediction of early-stage control.

The contributions of this work are fourfold:
\begin{itemize}
\item An interpretable bounded-rationality model that couples perceptual noise, looming aversion, and action delay directly to closed-loop control for the seconds following handover.
\item A sequential inference method that updates latent cognitive parameters from recent observations to maintain predictive accuracy as cognition changes.
\item A vehicle-in-the-loop study (n=41) demonstrating that the adaptive model improves short-horizon prediction over a non-adaptive baseline and anticipates degraded control earlier, supporting earlier and fewer-but-better interventions.
\item A physiological consistency analysis showing that inferred cognitive parameter fluctuations temporally align with real-time eye-tracking metrics, providing supporting evidence that the inferred parameters track cognition-related changes rather than merely serve to match observed trajectories.
\end{itemize}

\section{Related Work}

\subsection{Behavioral Prediction After Handover}
To mitigate risks during the handover period, researchers have explored models that predict driver behavior and control trajectories immediately after a takeover. Time-series and deep learning approaches dominate this category~\cite{chen2025systematic}. Recurrent neural networks (RNNs) and attention-based sequence models leverage rich inputs, including driver gaze, biometrics, and vehicle states, to forecast maneuvers or takeover outcomes~\cite{chen2025driver, bonyani2023dipnet, teshima2024determining}. Such data-driven models can achieve high accuracy when ample training data adequately captures the underlying scenario distribution. For instance, a deep multimodal network (DeepTake) classifies takeover intention, timing, and quality with up to 83–96\% accuracy~\cite{pakdamanian2021deeptake}. 

These methods demonstrate the potential of abundant sensory data and sequence modeling to anticipate drivers' initial control actions, yet they struggle under data scarcity and non-stationary driver states in the immediate post-handover moments. The first 0–3 seconds after a handover often present conditions unseen in training, as the driver's cognitive state is changing rapidly rather than at steady-state. One study restricted the prediction window to 3 seconds and still achieved only moderate F1-scores (64\% for "good vs. bad" control) despite using multiple physiological and contextual features~\cite{du2020predicting}. This indicates substantial uncertainty in these early-stage seconds that black-box models alone cannot fully resolve. 

Furthermore, most such models are trained offline and lack online adaptation; they do not update their parameters in real time as evidence accumulates. When confronted with an atypical driver response or a novel hazard in the first seconds, a static trained model may prove brittle and unreliable.

\subsection{Cognitive and Bounded-Rational Driver Models}
A parallel research stream has focused on models grounded in driver cognition and bounded rationality~\cite{janssen2024computational}. These works incorporate perceptual limitations, decision biases, and reaction delays into formal models of driver behavior. Examples include evidence accumulation models for driver reaction (e.g., braking decisions as drift-diffusion processes)~\cite{huang2025understanding, li2024human} and utility-based models that weight risks and rewards nonlinearly (prospect theory variant)~\cite{schmidt2014prospect, kong2025modeling}. Such models acknowledge that drivers do not always act optimally; instead, they satisfice under noise and time pressure~\cite{rendon2016effects}. By instantiating mechanisms like delayed perception of hazards, exaggerated risk aversion and underweighting of rare events, and resource-rational attention allocation~\cite{lieder2020resource}, these approaches yield more interpretable accounts of driver takeover behavior~\cite{wei2025active}. 

Cognitive models' parameters can be mapped to real cognitive states (e.g., high fatigue corresponds to higher inferred levels of uncertainty~\cite{sprajcer2023tired}), improving interpretability. However, they usually lack direct coupling to executable closed-loop control actions~\cite{engstrom2024resolving} and often assume static driver parameters during a takeover event. Moreover, these models are typically calibrated offline (e.g., using aggregate data or theoretical assumptions) and then used in simulation without online parameter updates. During the transient takeover phase, however, a driver's situational awareness and arousal are in a state of flux. A static model might mispredict if, for example, a driver suddenly snaps attention to a looming hazard, temporarily altering their "bounded" parameters. Without mechanisms for online learning or adaptation, cognitive models cannot fully account for within-event dynamics~\cite{engstrom2024resolving}.

\subsection{Positioning and Gap Statement}
Human–automation handover research establishes why those seconds are precarious, but the field relies on coarse proxies (reaction time, eyes-on-road) that only weakly predict control quality~\cite{soares2021takeover, mole2020predicting}. Behavioral sequence models can predict driver actions with impressive accuracy given sufficient representative data, yet they treat the driver as a static input-output mapping~\cite{zhu2023takeover} and cannot explain or adapt to the volatile cognitive state at takeover onset. Cognitive driver models, conversely, inject theoretical understanding of human limitations, but they typically remain disconnected from the actual vehicle control loop and do not update in real time as a driver regains capacity.

Our contribution addresses this gap by tightly coupling concrete cognitive mechanisms to a predictive driver-control model in a closed-loop fashion, while adapting this model online during the transient takeover phase. In doing so, we aim to predict early-stage control quality, not just reaction time or task outcome, within the critical first seconds after handover. To the best of our knowledge, no prior work has simultaneously combined specific, interpretable cognitive modeling with executable control signals and real-time parameter adaptation for this early post-handover interval, nor evaluated whether inferred cognitive states covary with time-resolved physiological markers. This approach directly addresses the identified gaps: it goes beyond surface behavior to the driver's internal state, links that state to control actions, continuously refines predictions as the takeover unfolds, and anchors these predictions in physiologically grounded cognitive fluctuations.

\section{Cognition-to-Control Coupling under Bounded Rationality}

We develop a cognition-integrated deep reinforcement learning framework to model driver behavior in the seconds following control handover under dynamic traffic conditions. The framework retains goal-directed optimality through its reinforcement learning objective, while explicitly instantiating bounded rationality.
To embed bounded rationality into the decision process, we construct mechanism-specific cognitive modules that represent perceptual noise, looming sensitivity, and reaction delay. These modules are integrated into the Partially Observable Markov Decision Process (POMDP) through three cognitive interfaces for observation, reward, and action. In addition, the cognitive parameters are directly incorporated into the policy network, enabling cognition to modulate how actions are produced. As a result, these cognitively influenced actions are ultimately realized as vehicle-level motion trajectories.

\subsection{Early Takeover Driving Process}

\begin{figure}[h]
  \centering
  \includegraphics[width=\linewidth]{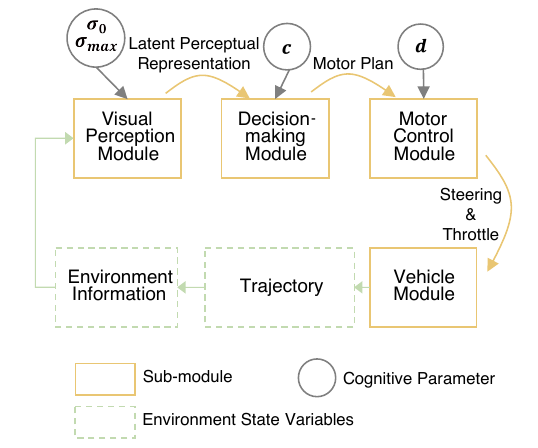}
  \caption{Schematic cognition-to-control coupling framework for early takeover driving.}
  \Description{A block-diagram cognition-to-control coupling framework for early takeover driving. Environmental state information enters a visual-perception module whose uncertainty is parameterized by sigma_0 and sigma_max; a decision-making module weights risk using looming-aversion parameter c; a motor-control module executes actions with delay parameter d; and a vehicle/environment module closes the feedback loop back to perception.}
  \label{fig:pomdp}
\end{figure}

The model simulates the driver's behavior after takeover through four interconnected submodules (Figure \ref{fig:pomdp}). It begins with the visual perception module, which processes incoming environmental data influenced by cognitive parameters, $\sigma_0$ and $\sigma_{\max}$ that modulate perceptual uncertainty. This module produces a latent representation of the environment, which feeds into the decision-making module, where the driver formulates a plan based on the perceived information. The looming-aversion parameter $c$ affects this decision-making by shaping how impending risks are weighted. The plan is then sent to the motor control module, which translates it into physical actions, accounting for action delay, parameterized by $d$. Finally, the vehicle module converts the driver's actions into vehicle movements, influencing the environment and creating feedback that is fed back into the visual perception module, closing the loop.

\subsection{Cognitive Mechanisms under Bounded Rationality}
During the post-handover interval, drivers must stabilize lateral and longitudinal motion under time pressure while a cascade of cognitive and sensorimotor factors shapes both scene interpretation and the closed-loop interaction between vehicle and environment. Residual effects from drivers' pre-handover behavioral state—e.g., engagement in non-driving-related tasks—preclude consistently optimal performance. Accordingly, we structure our modeling along a perception–cognition–motor pipeline that captures perceptual uncertainty, looming-averse risk appraisal, and delayed control execution.

\subsubsection{Perceptual Uncertainty}
Converging evidence from vision science indicates that sensory inputs to the human visual system are inherently noisy~\cite{faisal2008noise, kwon2015unifying}; likewise, observation noise is routinely incorporated in pedestrian behavior models to capture variability in decision making and locomotion~\cite{markkula2023explaining, wang2025pedestrian, wang2025modeling}. Guided by these findings, we inject range-dependent Gaussian perturbations into the observation channel to model perceptual uncertainty during the post-handover period. This noise-injection scheme emulates the post-handover period in which drivers re-establish gaze and situational awareness, leading to uncertainty and consequent bias in estimates of target position.

Based on the observation that human perceptual errors are approximately normally distributed~\cite{anderson2014auditory}, this paper further assumes that the agent's perception of range to surrounding vehicles is corrupted by zero-mean Gaussian noise.
Let $r_k$ denote the true ego-relative range to the other vehicle at step $k$.
The observation $\tilde{r}_k$ is modeled as:
\begin{equation}
  \tilde{r}_k \;=\; r_k \;+\; \varepsilon_k,
  \qquad \varepsilon_k \sim \mathcal{N}\!\bigl(0,\ \sigma_x^2(r_k)\bigr).
  \label{eq:meas_model}
\end{equation}

To reflect human distance discrimination, we assume the distance standard deviation grows approximately linearly with range, as supported by the concept of scalar variability~\cite{petzschner2015bayesian}, which posits that the standard deviation of magnitude estimates increases linearly with the mean of the stimulus.  
For numerical stability and identifiability, we adopt a Weber--Fechner-style saturation:
\begin{equation}
  \sigma_x(r) \;=\; \min\!\Bigl(\sqrt{\sigma_0^2 + \bigl(k_\sigma\, r\bigr)^2},\ \sigma_{\max}\Bigr),
  \label{eq:weber}
\end{equation}
where $\sigma_0$ is the near-range baseline noise, $k_\sigma$ controls the slope with distance, and $\sigma_{\max}$ caps far-range saturation.

Unless otherwise stated, we set $r_{\text{far}}=150~\text{m}$ and choose the slope $k_\sigma$ so that
$\sigma_x(r_{\text{far}})=\sigma_{\max}$ in \eqref{eq:weber}:
\begin{equation*}
  k_\sigma \;\triangleq\; \frac{\sqrt{\sigma_{\max}^2 - \sigma_0^2}}{\,r_{\text{far}}\,}.
\end{equation*}

Grounded in Bayesian perception~\cite{knill2004bayesian}, we estimate the distance of ego to surrounding vehicles with a Kalman filter that performs reliability-weighted temporal fusion of noisy measurements, maintaining a posterior over distance rather than a single point estimate. This operationalizes a context-dependent allocation of “trust” to sensory evidence—an algorithmic analogue of how the brain integrates uncertain observations over time.

\subsubsection{Looming Aversion}
Looming aversion denotes a robust, reflex-like bias to treat rapidly expanding objects in the visual field as urgent threats~\cite{delucia2008critical}. Evidence shows this cue materially shapes crossing decisions in pedestrian models~\cite{tian2022explaining}. In driving scenarios, the rapid increase in the volume of objects ahead often implies the risk of rear-end collisions. Accordingly, we incorporate a looming-weighted risk appraisal in our model so that quickly approaching hazards receive disproportionate influence on near-term control.

Mathematically, the driver’s looming cue for the lead vehicle is expressed via the inverse $\tau$—the ratio of a vehicle’s optical expansion rate to its size on the observer’s retina—which serves as an estimate of the inverse time-to-arrival (TTA)~\cite{markkula2016farewell, wang2025modeling}. 
The looming aversion is incorporated into the POMDP reward function as:
\begin{equation}
  R_{\text{looming}} \;=\;
  \begin{cases}
    - c\,\tanh\!\big({\hat{\tau}}^{-1}\big), & v>0,\\[2pt]
    0, & \text{otherwise}
  \end{cases},
  \label{eq:looming}
\end{equation}
where $c$ is the weighting coefficient for the looming aversion, $v$ denotes the closing speed, and $\hat{\tau}$ is the estimated TTA.

\subsubsection{Action Delay}
Human drivers do not act instantaneously, and the perception–decision–action pipeline introduces a non-negligible sensorimotor latency~\cite{kosinski2008literature, kuang2015does}. 
To capture this lag, we introduce an action delay before actions are executed on the vehicle.
In our discrete-time setting with step $\Delta t$, the delay is parameterized as
\begin{equation*}
    t_d = d \, \Delta t,
\end{equation*}
so a control computed at time $t-t_d$ is executed at time $t$, that is,
\begin{equation}
    a_t^{\text{exec}} = a_{t-t_d},
    \label{eq:actuation_delay_exec}
\end{equation}
where $a_t^{\text{exec}}$ denotes the control applied to the vehicle at time $t$.

\subsection{ POMDP with Embedded Cognitive Constraints}
\begin{figure}[h]
  \centering
  \includegraphics[width=\linewidth]{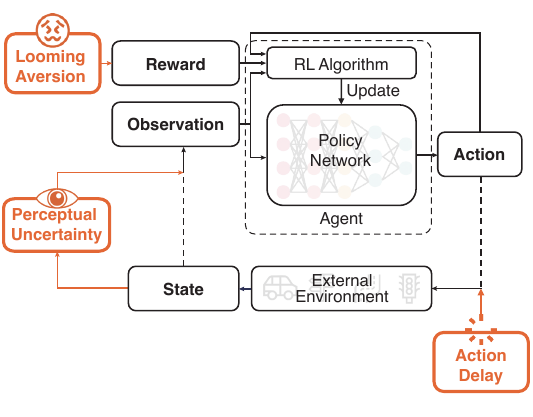}
  \caption{Reinforcement learning framework with embedded cognitive constraints for bounded-rationality modeling of driver behavior in post-takeover processes. The cognitive modules interface with specific components of the process: \textit{Perceptual Uncertainty} injects noise into the information transfer from environmental \textit{State} to the agent's \textit{Observation}; \textit{Action Delay} introduces latency in the feedback from \textit{Action} to the environment; and \textit{Looming Aversion} modulates the \textit{Reward} signal to reflect the influence of risk perception on the decision-making process in the face of potential risks.}
  \Description{A reinforcement-learning/POMDP framework diagram with embedded cognitive constraints. Perceptual-uncertainty injects noise from environment state to agent observation, action-delay introduces latency between chosen actions and their effect on the environment, and looming-aversion modulates the reward signal to reflect risk appraisal; these cognition modules interface with the standard state-observation-policy-action-environment loop.}
  \label{fig:rl_framework}
\end{figure}

The driver's dynamics in the post-takeover phase are formulated as a POMDP, where the previously introduced cognitive mechanisms are integrated into the structure, as illustrated in Figure \ref{fig:rl_framework}.

In this framework, the driver, modeled as the agent, relies on partial and noisy observations. The POMDP is defined as a tuple $\langle \mathcal{S}, \mathcal{A}, \Omega, T, R, O, \rho_0, \gamma \rangle$, where $\mathcal{S}$ is the state space, $\mathcal{A}$ is the action space, $\Omega$ is the observation space, $T(s'|s,a)$ is the state-transition distribution, $R(s,a)$ is the reward distribution, $O(o|s')$ is the observation distribution, $\rho_0(s)$ is the initial state distribution, and $\gamma \in [0,1)$ is the discount factor.

\subsubsection{State}
The simulation is conducted using the MetaDrive simulator~\cite{li2022metadrive}, which provides a comprehensive physics engine and realistic driving environment. At each time step $t$, the full environment is completely described by the state $s_t \in \mathcal{S}$, which is represented as
$s_t = \big[s_t^{\text{ego}},\, s_t^{\text{traffic}},\, s^{\text{map}},\, \theta_t\big]$,
where $s_t^{\text{ego}}$ collects the ego vehicle's dynamic variables (position $(x,y,z)$, velocity $(v_x,v_y,v_z)$, heading $\phi$, angular velocity $\omega$, steering angle $\delta$, throttle $a_{\text{throttle}}$, and brake $a_{\text{brake}}$); $s_t^{\text{traffic}}$ summarizes neighboring vehicles' states (e.g., IDs, positions, velocities, and headings); $s^{\text{map}}$ encodes the road context (lane geometry, intersections, traffic lights, and static obstacles); and $\theta_t = (\sigma_0, \sigma_{\max}, c, d)$ denotes the cognitive parameter state of the agent.

\subsubsection{Action}
At each time step, the agent executes an action $a_t \in \mathcal{A}$. The action is defined as a two-dimensional continuous vector $a_t = [u_t^{\text{steer}}, u_t^{\text{long}}]$, where:
\begin{itemize}
    \item $u_t^{\text{steer}} \in [-1, 1]$ represents the steering control: $-1$ indicates maximum left turn, $0$ indicates straight driving, and $+1$ indicates maximum right turn.
    \item $u_t^{\text{long}} \in [-1, 1]$ represents the longitudinal control (throttle/brake): $+1$ indicates maximum acceleration, $0$ maintains the current state, and $-1$ indicates maximum braking.
\end{itemize}

Action values are normalized control commands, scaled by MetaDrive's dynamics engine to fit vehicle limits. The simulator updates at 50 Hz, with a decision-making frequency of 10 Hz.

Human sensorimotor latency is modeled with a first-in-first-out (FIFO) action buffer, where control actions are executed after a delay of $t_d$ simulation steps, as determined by Equation \eqref{eq:actuation_delay_exec}.

\subsubsection{Transition}
The state transition follows the distribution $T(s'|s,a)$, which defines the probability of moving to state $s'$ from state $s$ with action $a$. In this framework, the MetaDrive physics engine handles vehicle motion and collisions. The transition dynamics are given by:
\begin{equation*}
s_{t+1} \sim T(\cdot|s_t, a_t),
\end{equation*}
where $T$ is defined by the Bullet-based vehicle dynamics in MetaDrive. Actions are executed at 50 Hz, updating the state at each step.

We do not alter MetaDrive's physical dynamics, ensuring consistency with standard simulation benchmarks and reproducibility.

\subsubsection{Reward}
To model human-like driving behavior that balances efficiency, safety, and comfort, the overall reward is constructed as a combination of \emph{dense}, \emph{sparse}, and \emph{cognitive} components. At each time step $t$, the instantaneous reward is defined as:
\begin{equation*}
    R(s_t, a_t) = R_{\text{sparse}}(s_t, a_t) + R_{\text{dense}}(s_t, a_t) + R_{\text{cognitive}}(s_t, a_t).
\end{equation*}

Each component term is detailed in the following paragraphs.

\textbf{Sparse Rewards.} Sparse rewards are only triggered at episode termination and capture successful completion or major failures:
\begin{equation*}
    R_{\text{sparse}} = R_{\text{success}} + R_{\text{road}} + R_{\text{crash}}.
\end{equation*}

\begin{enumerate}
    \item Success Reward ($R_{\text{success}}$): a positive reward for reaching the destination, reflecting the human driver's natural goal of reaching their destination.
    \item Out-of-Road Penalty ($R_{\text{road}}$): a negative reward for deviating from the drivable area, mimicking the need to stay within lanes for safety.
    \item Crash Penalty ($R_{\text{crash}}$): a negative reward for collisions, reflecting the goal to avoid accidents.
\end{enumerate}

\textbf{Dense Rewards.} Dense rewards are provided at each time step (10~Hz) to deliver continuous feedback during driving, reflecting the fact that human drivers continuously adjust their control based on ongoing progress, lane keeping, and speed regulation.
\begin{equation*}
    R_{\text{dense}} = R_{\text{driving}} + R_{\text{speed-control}}.
\end{equation*}

\begin{enumerate}
  \item Driving Reward ($R_{\text{driving}}$): Encourages forward progress and lateral stability, reflecting the driver's tendency to stay centered in the lane while moving toward the destination.
  
  \begin{equation*}
    R_{\text{driving}}
    = \alpha_{\text{drive}} \,
      \Delta l_t \,
      f_{\text{lat}}(t) \,
      I_{\text{road}}(t),
  \end{equation*}
  where $\alpha_{\text{drive}}$ is the driving reward coefficient, 
  $\Delta l_t = l_t - l_{t-1}$ is the longitudinal progress, and
  $I_{\text{road}}(t)\in\{0,1\}$ indicates whether the vehicle remains in the drivable area.
  
  The lateral stability factor $f_{\text{lat}}(t)$ is defined as:
  \begin{equation*}
    f_{\text{lat}}(t) = \mathrm{clip}\!\left(
        1 - 2 \left|\dfrac{d_t}{w_{\text{lane}}}\right|,\, 0,\,1
    \right),
  \end{equation*}
  where $d_t$ is the lateral deviation from the lane center, 
  and $w_{\text{lane}}$ is the lane width.

  \item Speed Control Reward ($R_{\text{speed-control}}$): This reward encourages the agent to maintain a speed near the target limit, penalizing both speeding and driving too slowly, reflecting human drivers' preference for efficient and safe driving.

\end{enumerate}

\begin{equation*}
  R_{\text{speed-control}} = R_{\text{track}} + R_{\text{wall}} + R_{\text{behavior}}.
\end{equation*}

\begin{itemize}
  \item $R_{\text{track}}$: Tracking Reward encourages the agent to maintain the target speed using a smooth Huber loss: 
  $R_{\text{track}} = -k_{\text{track}} \cdot \mathcal{H}(\Delta v_t, \delta_v)$,
  where $k_{\text{track}}$ is the tracking reward coefficient, $\Delta v_t$ is the speed error (the difference between the current speed $v_t$ and the target speed $v_{\text{target}}$), 
  and $\mathcal{H}(\cdot)$ denotes the standard Huber loss with threshold $\delta_v$.

  \item $R_{\text{wall}}$: Soft Wall Reward applies a smooth penalty only when the vehicle exceeds the target speed:
  $R_{\text{wall}} = -\kappa \cdot I_{\text{over}}(t) \cdot \left[\text{softplus}(\Delta v_t)\right]^2 $,
  where $\kappa$ is the soft wall penalty coefficient, $\text{softplus}(x)$ is a smooth approximation to $\max(0, x)$, and $I_{\text{over}}(t) \in \{0,1\}$ indicates whether the vehicle is exceeding the target speed.

  \item $R_{\text{behavior}}$: Behavior Guidance Reward encourages appropriate deceleration actions when speeding:
  $R_{\text{behavior}} = I_{\text{over}}(t) \cdot \left(\mu \cdot acc_t^- - \nu \cdot acc_t^+\right)$,
  where $\mu$ and $\nu$ are the deceleration reward and acceleration penalty coefficients respectively, 
  $acc_t = (v_t - v_{t-1})/\Delta t$ is the estimated acceleration, 
  $acc_t^+ = \max(acc_t, 0)$ captures additional acceleration (which is penalized when the vehicle is already speeding), and 
  $acc_t^- = \max(-acc_t, 0)$ captures deceleration (which is rewarded when the vehicle is speeding).

\end{itemize}

\textbf{Cognitive Reward.} To model bounded-rational behavior, we introduce a cognitive component that captures the driver's looming-averse risk appraisal. 
This term is provided at each time step (10~Hz) and penalizes situations in which the looming aversion becomes excessively high, as defined in Equation \eqref{eq:looming}.
\begin{equation*}
  R_{\text{cognitive}} = R_{\text{looming}}.
\end{equation*}

\textbf{Reward Coefficients.}
The reward function structure is summarized in Table~\ref{tab:reward_coeffs}. The coefficients balance progress, safety, and cognitive realism, and remain consistent across all experiments unless noted.
In $R_{\text{looming}}$, the coefficient $c$ varies as a cognitive parameter.

\begin{table}[t]
  \centering
  \caption{Reward-related terms and parameters used in training. Positive values indicate rewards, and negative values indicate penalties where applicable.}
  \label{tab:reward_coeffs}
  \begin{tabular}{l l c l}
  \toprule
  \textbf{Category} & \textbf{Symbol} & \textbf{Value} & \textbf{Unit} \\
  \midrule
  \multirow{3}{*}{Reward term} & $R_{\text{success}}$ & $+100$ & -- \\
   & $R_{\text{road}}$ & $-8$ & -- \\
   & $R_{\text{crash}}$ & $-8$ & -- \\
  \midrule
  \multirow{6}{*}{Parameter} & $\alpha_{\text{drive}}$ & 0.4 & -- \\
   & $k_{\text{track}}$ & 0.12 & -- \\
   & $\kappa$ & 0.15 & -- \\
   & $\mu$ & 0.3 & -- \\
   & $\nu$ & 0.2 & -- \\
   & $\delta_v$ & 1 & m/s \\
  \midrule
  \multirow{1}{*}{Others} & $v_{\text{target}}$ & 27.8 & m/s \\
  \bottomrule
  \end{tabular}
\end{table}

\subsubsection{Observation}
The agent does not perceive the full state $s_t$ directly but receives an observation $o_t$ according to the observation distribution $O(o|s')$, which models the probability of receiving an observation $o$ given the next state $s'$. In our framework, the observation is generated as:
\begin{equation*}
o_t \sim O(\cdot|s_{t+1}).
\end{equation*}

The agent is assumed to have near-precise knowledge of its own motion state $s_t^{\text{ego}}$ (e.g., position, heading, and velocity), which is consistent with common practices in continuous control reinforcement learning and the availability of self-vehicle perception (e.g., odometry and lane detection) in modern driving platforms. However, its perception of surrounding traffic, denoted as $s_t^{\text{traffic}}$, is imperfect due to perception noise and is interpreted through its own cognitive understanding. 

The perception noise is modeled using the Gaussian noise framework described in Equation \eqref{eq:meas_model}, where the observed traffic state $\tilde{s}_t^{\text{traffic}}$ is obtained by adding distance-dependent Gaussian noise to the true traffic state:
\begin{equation*}
\tilde{s}_t^{\text{traffic}} = s_t^{\text{traffic}} + \mathcal{N}(0, \sigma_x^2(r)),
\end{equation*}
where $r$ denotes ego-relative range and $\sigma_x(r)$ represents the range-dependent noise standard deviation as defined in Equation \eqref{eq:weber}.

Therefore, the observation available to the agent is $o_t = [s_t^{\text{ego}}, \hat{s}_t^{\text{traffic}}, s^{\text{map}}]$, where $\hat{s}_t^{\text{traffic}}$ represents the Kalman-filtered estimate of the traffic state.

The cognitive state $\theta_t$ is an internal variable of the agent and not a direct observation from the environment. Since different parameters ($\sigma_0, \sigma_{\max}, c, d$) influence the state-action process, they are fed as inputs to the reinforcement learning (RL) policy network, thereby conditioning the RL on these parameters~\cite{moon2022speeding, li2023modeling}. These parameters are referred to as non-policy parameters to distinguish them from the network's own parameters, such as connection weights and biases.

\subsection{Reinforcement Learning and Cognition-Aware Policy Modulation}

\subsubsection{Algorithm and Network}
We use proximal policy optimization (PPO)~\cite{schulman2017proximal}, an on-policy actor–critic method with a clipped surrogate objective, to simulate drivers' goal-directed control policies. PPO balances sample efficiency and simplicity, and is widely validated on continuous-control tasks~\cite{ wang2020truly, zhang2022proximal}. In our setup, the policy is trained in closed loop using MetaDrive within the Stable-Baselines3 (SB3) framework~\cite{raffin2021stable}, with a discrete time step of $\Delta t = 0.1~\mathrm{s}$.

The architecture uses a decoupled actor–critic for continuous control. The input is a 283-dimensional vector, including 275 MetaDrive observations, 4 cognitive parameters, and 4 masks. Both actor and critic are two-layer MLPs with 256 hidden units and Tanh activations. The actor outputs a bivariate Gaussian for steering and longitudinal commands, and the critic provides a scalar state value.

\subsubsection{Cognitive Params Integration}
We append cognitive (non-policy) parameters directly to the RL input and, unless noted otherwise, draw them from uniform priors and refresh them within an episode on a fixed resampling interval $H_{\mathrm{resample}}$. 
Specifically, every $H_{\mathrm{resample}}=5$ control steps, we resample the looming-aversion weight $c\!\sim\!\mathcal{U}[0.0,\,10.0]$, the near-range noise floor $\sigma_{0}\!\sim\!\mathcal{U}[0.0,\,1.0]~\mathrm{m}$, the far-range saturation $\sigma_{\max}\!\sim\!\mathcal{U}[0.0,\,5.0]~\mathrm{m}$, and the discrete action delay $d\!\in\![0,20]$, with the constraint $\sigma_{0} \le \sigma_{\max}$.
Parameters are held piecewise-constant between refreshes, which facilitates adaptive parameter inference and short-horizon prediction of driver behavior.

\subsubsection{Training Details}
We train our agent on procedurally generated straight-highway scenarios through two phases.

The first phase is cognition-off pretraining to disable cognition by masking the cognitive channels so that the network cannot use them, yielding a policy that approximates an unbounded-rational driver. 
The agent is trained for $5\times10^{7}$ environment steps under a curriculum schedule that automatically increases traffic density as training progresses. Concretely, the curriculum consists of three stages: (1) Stage 1 spans the first 10\% of training, where density rises from 0.03 to 0.06; (2) Stage 2 covers the next 10\% of training, further increasing density from 0.06 to 0.09; and (3) Stage 3 accounts for 20\%-100\% of training, during which density continues to grow from 0.09 toward 0.12 as progress advances. This progressive design stabilizes early convergence while gradually exposing the policy to more dynamic interactions.

The second phase is cognition-on finetuning to enable the cognitive parameters and continue training for another $5\times10^{7}$ steps, holding the traffic density in the $0.06$--$0.10$ range. This setup stresses cognition-induced variability (noise, looming sensitivity, action delay) over extreme traffic densities, so the policy learns cognition-aware modulation during closed-loop rollouts.

\section{Online Adaptation of Latent Cognitive State for Action Prediction}

Building upon the theoretical framework established in the previous sections, this section presents the implementation of our online cognitive state inference system for real-time prediction of driver post-handover behavior. The system operates in two interconnected phases: first, it dynamically infers latent cognitive parameters from observed driving trajectories using the particle filter framework~\cite{schwarting2019social}; second, it leverages these inferred parameters to predict driver behavior after control handover in an adaptive manner.

The key innovation lies in the closed-loop integration of cognitive parameter estimation and behavior prediction, enabling the system to continuously adapt predicted behavior in response to changes in driver cognitive state.

\subsection{Dynamical Inference of Latent Cognitive Parameters}
This section presents a particle-filter--based framework for dynamic inference of latent cognitive (non-policy) parameters, enabling Bayesian sequential estimation of a multi-dimensional cognitive state. The objective is to online identify the time-varying driver cognition vector $\boldsymbol{\theta}_t = [\sigma_0,\ \sigma_{\max},\ c,\ d]$ from the observed post-handover vehicle trajectory sequence $\{y_t\}_{t=1}^{T}$, capturing fluctuations in cognition over time.

To realize online adaptation within our closed-loop framework, we extend classical Bayesian filtering from a single measurement to a sliding window of $L$ consecutive observations. The overall idea is to use the past $L$ observations $y_{k-L:k}$ to find the cognitive parameters $\theta_{k-L}$ that best explain these observations, thereby aligning inference with control. Given the proposed post-handover closed-loop simulator that can produce the windowed predicted trajectory $\hat y_{k-L:k}(\theta_{k-L})$, the measurement function is written as
\begin{align}
p\!\big(y_{k-L:k}\mid \theta_{k-L}\big)\ \propto\ 
\mathcal{N}\!\big(y_{k-L:k}\,;\ \hat y_{k-L:k}(\theta_{k-L}),\ \Sigma\big),
\label{eq:win_like_gauss}
\end{align}
where $\Sigma$ is a diagonal covariance. The predicted trajectory $\hat y_{k-L:k}(\theta_{k-L})$ is generated by rolling out the same closed-loop simulator used for behavior prediction, with its cognitive modules configured by $\theta_{k-L}$. This keeps the statistical measurement model consistent with the same cognitive modules used during control execution.

To make inference feasible in real time, when evaluating the measurement function only, we adopt a slow-variation approximation within the window, assuming the parameters remain approximately constant, $\theta_{k-L+1:k}\approx \theta_{k-L}$. Under this approximation, the window likelihood factors across time and becomes
\begin{align*}
p\!\big(y_{k-L:k}\mid \theta_{k-L}\big)\ \approx\ 
\prod_{s=k-L+1}^{k}\,
\mathcal{N}\!\big(y_s\,;\ \hat y_s(\theta_{k-L}),\ \Sigma\big).
\end{align*}

Within the standard prediction–update recursion of Bayesian filtering, the prior at the left endpoint $k-L$ is given by
\begin{align*}
p(\theta_{k-L}\mid y_{0:k-1})
\;=\;\int p(\theta_{k-L}\mid \theta_{k-L-1})\,
           p(\theta_{k-L-1}\mid y_{0:k-1})\, d\theta_{k-L-1},
\end{align*}
and our implementation instantiates the parameter dynamics with an additive Gaussian random walk:
\begin{align*}
% \theta_t \;=\; \theta_{t-1} + w_t,\qquad w_t\sim \mathcal{N}(0,Q)
% \;\;\Longrightarrow\;\;
p(\theta_t\mid \theta_{t-1}) \;=\; \mathcal{N}\!\big(\theta_t;\,\theta_{t-1},\,Q\big).
\end{align*}
where $Q$ is the process-noise covariance that controls how quickly the latent parameters are allowed to drift over time.

After receiving the new observation $y_k$, we update the prior using the observations in the window $[k-L+1,\,k]$ to obtain the posterior over the window-anchored parameters:
\begin{align}
p(\theta_{k-L}\mid y_{0:k})
\;=\;\frac{\,p\!\big(y_{k-L:k}\mid \theta_{k-L}\big)\;p(\theta_{k-L}\mid y_{0:k-1})\,}
           {\displaystyle \int p\!\big(y_{k-L:k}\mid \theta_{k-L}\big)\;p(\theta_{k-L}\mid y_{0:k-1})\, d\theta_{k-L}}.
\label{eq:posterior_window}
\end{align}

Combining the above elements yields a concrete online inference routine. We now describe two components that implement it: a prediction-based measurement module and a particle-filter update.

\subsubsection{Algorithm A: Prediction-Based Measurement.}
This module evaluates the windowed measurement term by rolling out the closed-loop simulator under a candidate parameter and comparing to observed trajectories. Given the window observations $y_{k-L:k}$ and a candidate parameter $\hat\theta_{k-L}$, this module rolls out the closed-loop simulator and, by comparing predictions to the observed trajectory, produces the window likelihood $p(y_{k-L:k}\mid \hat\theta_{k-L})$ as follows.
\begin{enumerate}
  \item \textbf{Closed-loop prediction:} Configure cognitive modules with $\hat\theta_{k-L}$ and roll the environment forward from the initial state extracted from $y_{k-L}$ for $L$ steps.
  \item \textbf{Likelihood computation:} The module evaluates the multivariate Gaussian likelihood from equation \eqref{eq:win_like_gauss}. This involves calculating the Mahalanobis distance between the observed trajectory and the simulated one, accounting for dimension-wise variance scaling encoded  in the covariance matrix $\Sigma$.
\end{enumerate}

\subsubsection{Algorithm B: Particle Filter Update}
This sequential Monte Carlo routine performs the prediction–update steps over a particle set, using the windowed likelihood from Algorithm A for weighting.
Starting from a particle set $\{\,\theta_{k-L-1}^{(i)},\,w_{k-L-1}^{(i)}\,\}_{i=1}^{N}$ and the window observations $ y_{k-L:k}$, this routine carries out prediction–update steps using the windowed likelihood from Algorithm A, and returns the posterior summary (mean and variance) together with the updated particle set $\{\,\theta_{k-L}^{(i)},\,w_{k-L}^{(i)}\,\}_{i=1}^{N}$.
\begin{enumerate}
  \item \textbf{Propagation:} Each particle's state is propagated forward according to the random walk dynamics:
  \begin{align*}
  \theta_{k-L}^{(i)} \;=\; \theta_{k-L-1}^{(i)} + \xi^{(i)},\qquad \xi^{(i)}\sim \mathcal{N}(0,Q).
  \end{align*}
  \item \textbf{Weighting:} The weight of each propagated particle is updated based on the geometric mean likelihood of the observation window, which implements the Bayesian posterior update from Equation \eqref{eq:posterior_window} in a particle filter framework. In log space, the weight update uses averaged log-likelihoods over the $L$ steps:
  \begin{align*}
  \log w_{k-L}^{(i)}
  \;=\; \log w_{k-L-1}^{(i)} + \frac{1}{L} &\sum_{s=k-L+1}^{k} \log p\!\big(y_s \mid \theta_{k-L}^{(i)}\big),\\
  \sum_{i=1}^{N} w_{k-L}^{(i)} \;&=\; 1.
  \end{align*}
  \item \textbf{Resampling:} To prevent particle degeneracy, systematic resampling is performed if the effective sample size (ESS), a measure of particle diversity, drops below a threshold, set to $N/2$ in our implementation.
  \begin{align*}
  \mathrm{ESS}_{k-L} \;=\; \frac{1}{\sum_{i=1}^{N}\big(w_{k-L}^{(i)}\big)^2}
  \end{align*}
  \item \textbf{Posterior Estimation:} The posterior distribution is summarized by computing the weighted mean and variance of the particles:
  \begin{align*}
    \hat\theta_{k-L} \;=\; \sum_{i=1}^{N} w_{k-L}^{(i)}\,\theta_{k-L}^{(i)},\qquad
    \mathrm{Var}\,\theta_{k-L} \;=\; \sum_{i=1}^{N} w_{k-L}^{(i)}\big(\theta_{k-L}^{(i)}-\hat\theta_{k-L}\big)^2.    
  \end{align*}
\end{enumerate}

\subsection{Online adaptive behavior prediction}

With the inferred cognitive parameters $\hat\theta_t$ obtained from the particle filter framework, we can now perform real-time prediction of the driver's post-handover behavior. This adaptive prediction module leverages the estimated cognitive state to forecast driver actions and responses over a finite horizon after control transition.

The prediction process operates as follows: given the current vehicle state $x_t$ and the latest cognitive parameter estimate $\hat\theta_t$, the simulator rolls out predicted driver trajectories $\hat y_{t+1:t+H}$ over a horizon $H$ using the inferred cognitive modules.

This approach enables personalized prediction of post-handover driving behavior that adapts to individual cognitive characteristics in real-time. As new observations become available, the cognitive parameters are continuously updated through the inference framework, allowing the prediction module to track changes in the driver's cognitive state and maintain accurate forecasting performance across varying handover scenarios.

\section{Study 1: Data collection}
To ground our computational model in real-world behavior, we collected high-fidelity measurements of driver responses in the first seconds after control handover—capturing the initial control dynamics when cognitive states shift rapidly and safety risk is elevated—immediately after drivers resumed manual control from automation.

\subsection{Participants and Apparatus}
We invited 41 participants between 20 and 35 years old (25 males, 16 females; ages: mean (M)=23.05 years, standard deviation (SD)=2.11) with diverse educational backgrounds from a local university. Participants reported a mean driving experience of 3.60 years and an average lifetime driving mileage of 33,520 km. Within the 24 hours prior to testing, they maintained regular sleep and rest and refrained from alcohol consumption and vigorous exercise to ensure alertness. The study was approved by the institutional ethics committee, and written informed consent was obtained from all participants.

\begin{figure*}[!t]
  \centering
  \includegraphics[width=\textwidth]{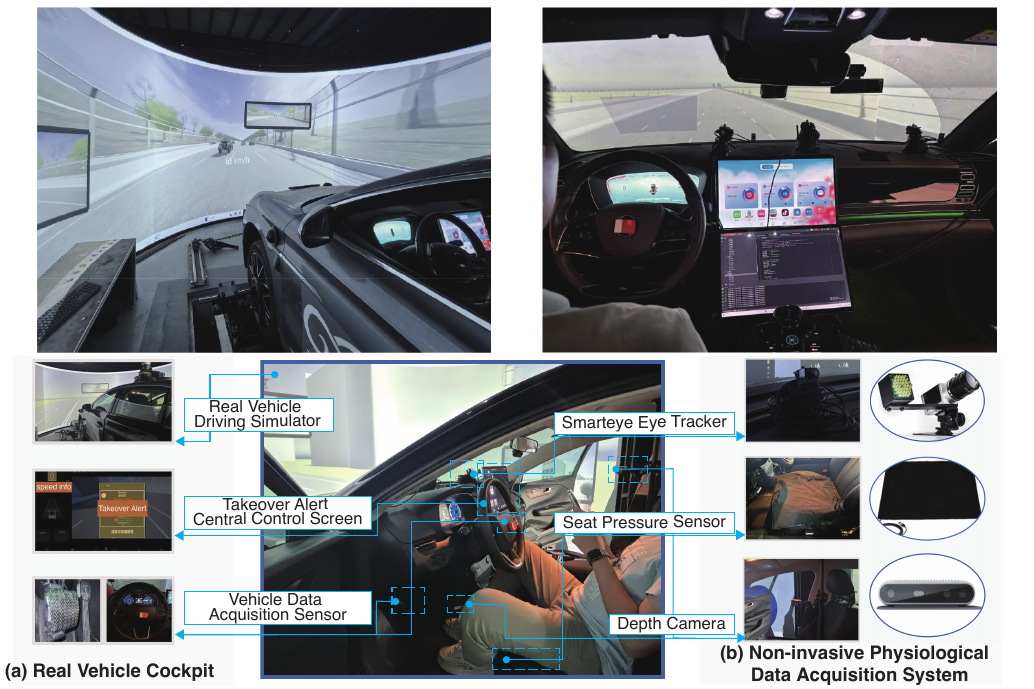}
  \caption{Takeover experiments on a high-fidelity vehicle-in-the-loop driving simulator.}
  \Description{Illustration/photo of a high-fidelity vehicle-in-the-loop takeover driving simulator setup. It includes a real vehicle cockpit, an immersive multi-projector wrap-around display, an HMI/touchscreen for takeover prompts, and instrumented driver inputs (steering wheel and accelerator/brake pedals) for recording steering and pedal actions.}
  \label{fig:simulator}
\end{figure*}

Experiments were conducted on a high-fidelity, full-scale driving simulator that reproduces autonomous-driving takeover interactions (Figure \ref{fig:simulator}). The simulator consists of a production BYD Han cockpit embedded in a wrap-around, edge-blended projection system using three high-definition (HD) projectors (cylindrical screen radius=2.8~m, aperture width=5~m, aggregate resolution=1080×5760), providing an immersive visual field with multi-channel perceptual coherence. A center-console touchscreen delivered takeover prompts. Driver inputs were instrumented with three inertial measurement unit (IMU)-based sensors: two serial-connected modules rigidly mounted on the accelerator and brake pedals, and a wireless module on the steering wheel. Sensor streams were transmitted over User Datagram Protocol (UDP) with low latency at 60 Hz to capture throttle, braking, and steering. The steering wheel incorporated programmable force-feedback with multi-level damping, enabling high-fidelity interaction with three control degrees of freedom. Throughout the experiment, participants remained seated in the real-vehicle cockpit. On the software side, a tightly coupled co-simulation stack integrated RoadRunner, CARLA (v0.9.15), and Unreal Engine 4 to unify virtual road modeling, real-time traffic scene rendering, and closed-loop interaction/control.

\subsection{Design}
\subsubsection{Takeover scenario}
\begin{figure}[h]
  \centering
  \includegraphics[width=\linewidth]{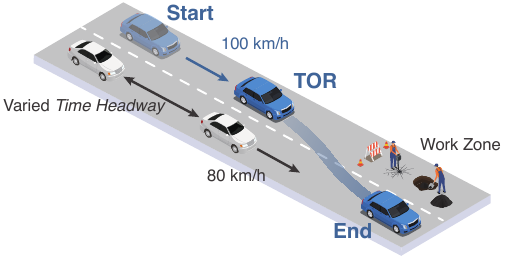}
  \caption{Experimental settings of the unexpected highway work-zone scenario. A takeover request (TOR) is issued as the ego vehicle approaches the upcoming work-zone obstruction, with background flow density manipulated via time headway (TH).}
  \Description{A schematic of the unexpected highway work-zone takeover scenario. On an eight-lane divided motorway, the ego vehicle approaches a work-zone obstruction occupying the leftmost lane; a takeover request (TOR) is issued as the ego nears the obstruction, and background traffic-flow density is manipulated via time headway (TH) (e.g., 1.75 s, 2.0 s, 2.25 s).}
  \label{fig:scenario_design}
\end{figure}

Informed by national reports on vehicle data, we instantiated a freeway “unexpected work-zone avoidance” scenario~\cite{boggs2020exploring} as shown in Figure \ref{fig:scenario_design}. On an eight-lane divided motorway (4×4), the ego vehicle traveled at 100 km/h and encountered a work zone occupying the leftmost lane; the driver needed to promptly take over to decelerate and execute a rightward lane change to bypass the obstruction. 
For traffic-flow parameterization, the nominal cruising speeds of the neighboring vehicles were set to 80 km/h, and flow density was manipulated via time headway (TH) at three levels: 2.25 s (low), 2.0 s (medium), and 1.75 s (high).

\subsubsection{Primary task}
During the simulated driving experiment, whenever a takeover was required, the system issued a takeover prompt. After re-establishing situational awareness, the driver executed the necessary evasive maneuver to complete the takeover. Depressing the brake pedal served as the takeover trigger: the automated driving system immediately disengaged, and full vehicle control (longitudinal and lateral) transferred to the driver.

\subsubsection{Initial Conditions for Scenario Diversity}

To create diverse initial takeover conditions within a unified framework, we designed scenarios along three dimensions: Non-Driving Task, Takeover Lead Time, and Takeover Request format. These dimensions captured intrinsic cognitive load, temporal resource availability, and cue presentation characteristics, jointly shaping drivers’ takeover cognition and behavioral responses.

\begin{enumerate}
\item \textbf{Non-Driving Task (NDRT).}
Guided by Multiple Resource Theory (MRT)~\cite{wickens2008multiple}, we defined four NDRT conditions with distinct resource profiles, each implemented as a concrete secondary task. For each experimental block, participants received specific instructions for the assigned NDRT:

\begin{enumerate}
\item Non-NDRT (monitoring the road): participants were instructed not to perform any secondary activity and to devote all cognitive resources to roadway surveillance.
\item NDRT (watching a tablet video): participants selected a preferred video from a set of preloaded videos on a tablet and were asked to prioritize watching it during automated driving when it was safe to do so; they could reallocate gaze to the roadway as needed. This task primarily engages visual and auditory channels with minimal manual interaction.
\item NDRT (playing the smartphone game "2048"): participants played the game 2048 on a smartphone and were asked to keep playing and try to maximize their score during automated driving when it was safe to do so; when needed, they could reallocate gaze to the roadway. This task requires dynamic allocation across visual and visuo-manual channels.
\item NDRT (reading a book): participants read a self-selected printed book and were asked to prioritize reading during automated driving when it was safe to do so; they could reallocate gaze to the roadway as needed. This task involves sustained visual attention and semantic processing.
\end{enumerate}

\item \textbf{Takeover Lead Time (TLT).} 
Consistent with prior evidence that each additional second of advance warning increases available takeover time by 0.2–0.3 s~\cite{gold2018modeling}, TLT was set to four equally spaced levels (4, 6, 8, and 10 s). This factor emulated varying risk-alert severities and tested how the warning window modulated cognitive processes during takeover—situational reconstruction, attentional reorientation, and control transition.

\item \textbf{Takeover Request (TOR).} 
Building on a baseline prompt (buzzer plus “Please take over” text) and prior demonstrations of the benefits of acoustic tuning and risk-information presentation~\cite{li2023human, wright2018effective}, we parameterized the TOR strategy along three orthogonal dimensions instead of describing it only qualitatively: 

\begin{enumerate}
  \item Acoustic modulation: the auditory cue was a pure-tone beep with two intensity levels. 
  The basic setting used a lower loudness and frequency (within 50–70 dB and 200–300 Hz, 
  presented once), whereas the urgent setting used a higher loudness and frequency 
  (within 80–90 dB and 400–500 Hz, presented twice with a short interval) to increase 
  perceived urgency.

  \item Modal configuration: the alert was delivered either as an auditory-only cue or as 
  an audiovisual cue, where the beep was synchronized with a high-contrast warning banner 
  on the central human--machine interface (HMI) displaying the takeover text in large bold font.

  \item Semantic enrichment: the message content was either a generic prompt 
  (“Please take over”) or a risk-informed prompt that added situational information, 
  for example: “Take over. Road construction ahead. Be aware of vehicles behind you.”
\end{enumerate}

Combining the two acoustic levels, two modality settings, and two semantic settings resulted in eight TOR configurations, ranging from a basic auditory-only alert to an urgent, risk-informed audiovisual alert, as summarized in Table~\ref{tab:tor_configurations}.

\begin{table*}[t]
\centering
\begin{tabular}{c l l l}
\toprule
\textbf{No.} & \textbf{Acoustic Modulation} & \textbf{Modal Configuration} & \textbf{Semantic Enrichment} \\
\midrule
1 & Low loudness, low frequency & Auditory-only & None \\
2 & High loudness, high frequency & Auditory-only & None \\
3 & Low loudness, low frequency & Audiovisual & None \\
4 & Low loudness, low frequency & Auditory-only & Risk-informed \\
5 & High loudness, high frequency & Audiovisual & None \\
6 & High loudness, high frequency & Auditory-only & Risk-informed \\
7 & Low loudness, low frequency & Audiovisual & Risk-informed \\
8 & High loudness, high frequency & Audiovisual & Risk-informed \\
\bottomrule
\end{tabular}
\caption{TOR Configuration Descriptions Based on Acoustic Modulation, Modality, and Semantic Enrichment}
\label{tab:tor_configurations}
\end{table*}

\end{enumerate}

\subsection{Procedure and Measurements}

We adopted a mixed-level orthogonal design to accommodate unequal factor cardinalities. Specifically, we used a mixed-level orthogonal array denoted $L_{32}(8^1 \cdot 4^2)$, where $L_{32}$ denotes 32 experimental runs and $8^1 \cdot 4^2$ denotes one eight-level factor and two four-level factors. This yielded a three-factor, multi-level design in which the factors were the TOR strategy with 8 levels, TLT with 4 levels, and NDRT with 4 levels. The orthogonal array sampled a level-balanced set of 32 combinations from the full $8 \times 4 \times 4 = 128$ possible TOR--TLT--NDRT configurations, enabling statistically efficient and unbiased estimation of the main effects of each factor. The resulting 32 TOR--TLT--NDRT combinations were randomly embedded into distinct takeover sub-scenarios, and, for each participant, a Latin-square counterbalancing scheme was used to determine trial order and mitigate sequence and learning effects.

\begin{figure*}[t]
  \centering
  \includegraphics[width=\textwidth]{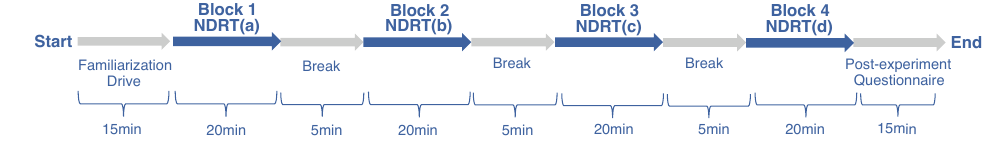}
  \caption{Complete experimental procedure for each participant.}
  \Description{A flowchart/timeline of the complete experimental procedure for each participant, including informed consent, seat adjustment and device calibration, a practice drive, four experimental blocks (with breaks) covering different non-driving task conditions, and a post-experiment questionnaire; participants complete multiple takeover trials across the session.}
  \label{fig:procedure}
\end{figure*}

To manage session length and cognitive load, the experiment was partitioned into four 20-min blocks by NDRT type, separated by 5-min breaks, preserving the independence of manipulations and cross-session comparability. Each participant completed 32 takeover trials in total (41 participants $\times$ 32 trials = 1312 takeover episodes), in addition to a 15-min familiarization/practice drive. The procedural workflow comprised informed consent, seat adjustment, device calibration, the familiarization/practice drive, four experimental blocks, and a 15-min post-experiment questionnaire, so that each participant spent approximately 125 min as shown in Figure \ref{fig:procedure}.

\section{Study 2: Model Evaluation}
In this study, we evaluate the proposed adaptive behavior model using post-handover driving data collected in Study~1, focusing on the interval from takeover onset until either task completion or collision. 

Unless otherwise stated, all rolling inference and prediction in Study~2 uses a sliding window of $L=5$ steps (0.5~s) and a look-ahead horizon of $H=30$ steps (3~s). For particle-filter inference, particles are initialized via Latin hypercube sampling with $N=20$ particles, and the search ranges are $\sigma_0 \in [0.0, 1.0]$, $\sigma_{\max} \in [0.0, 5.0]$, $c \in [0.0, 10.0]$, and $d \in [0, 20]$, with the constraint $\sigma_{0} \le \sigma_{\max}$.

Our evaluation proceeds at two levels. First, we conduct a quantitative assessment that characterizes overall model reliability, early-warning performance, lead-time coverage, and the consistency between inferred cognitive parameters and observed behavior across a broad set of takeover episodes. Second, we present a qualitative, mechanism-oriented analysis of a representative takeover episode, tracking prediction accuracy, the evolving collision-risk timeline, and the coupled trajectories of cognitive parameters and driver control actions to illustrate how the model anticipates and interprets collision risk.

\subsection{Quantitative Evaluation 1: Early Warning and Lead-Time Coverage}
To assess the model’s effectiveness in broader settings, we randomly sampled 200 collision-containing takeover episodes from our dataset, corresponding to approximately 57.3\% of all collision cases, due to the high computational cost of rolling-horizon evaluation. Each method performs rolling prediction with a window size $L=5$ (0.5~s) and a look-ahead horizon $H=30$ steps (3~s), outputting at every step a decision on whether a collision will occur within the horizon. We compared against two baselines:
(1) Constant-Velocity (CV): all vehicles are assumed to maintain the current speed and heading throughout the prediction window;
(2) Vanilla PPO: no cognitive parameters, the policy network approximates a "rational" driver.

\subsubsection{Early-Warning Hit Rate}
Early warning performance was evaluated by checking whether models predicted collisions at least 0.5 seconds before they occurred. A warning was considered successful if $t_{\text{col}} - t_{\text{flag}} \geq 0.5$ seconds.
Here, $t_{\text{flag}}$ denotes the time when the model first issues a collision warning, and $t_{\text{col}}$ denotes the ground-truth collision time.

The results showed significant differences in predictive capability. The cognition-adaptive approach achieved early warnings in 89.5\% of the 200 collision episodes, outperforming Vanilla PPO (41.5\%) and Constant-Velocity (22.0\%). This pattern indicates the advantage of incorporating dynamic cognitive parameters that capture evolving human factors during takeovers.

CV struggles to predict changes in motion due to acceleration, steering, and lane changes. Vanilla PPO fits average behavior but misses non-stationary human factors, often failing when risk is just beginning to show. In contrast, the cognition-adaptive method updates hidden variables in real-time, detecting collision risks early and providing much higher early-warning coverage.

\subsubsection{Lead-Time Coverage}

\begin{table}[h]
\centering
\caption{Collision warning coverage rates at different lead-time thresholds}
\label{tab:lead_time}
\begin{tabular}{lccc}
\toprule
\textbf{Method} & \textbf{$\geq$0.5 s} & \textbf{$\geq$1 s} & \textbf{$\geq$2 s} \\
\midrule
Cognition-Adaptive & 89.5\% & 80.5\% & 58.5\% \\
CV & 22.0\% & 19.5\% & 15.0\% \\
Vanilla PPO & 41.5\% & 12.0\% & 0.0\% \\
\bottomrule
\end{tabular}
\end{table}

Coverage is reported at three lead-time thresholds: $\geq$0.5~s, $\geq$1~s, and $\geq$2~s (Table \ref{tab:lead_time}). The cognition-adaptive model achieved 89.5\%, 80.5\%, and 58.5\% at these thresholds, respectively, outperforming CV (22.0\%, 19.5\%, 15.0\%) and Vanilla PPO (41.5\%, 12.0\%, 0.0\%).

At the critical 1-second threshold, the cognition-adaptive model maintained a coverage of 80.5\%, while the Vanilla PPO dropped to 12.0\%. At 2 seconds, only the cognition-adaptive model maintained meaningful predictive capability (58.5\%).

These results reflect the model assumptions: CV's coverage is low and slowly degrades, while Vanilla PPO’s performance collapses beyond 0.5~s, failing to anticipate non-stationary human factors. The cognition-adaptive model retains strong performance even at longer lead times, highlighting its ability to model and adapt to human cognitive limitations and extend the actionable warning horizon.

\subsection{Quantitative Evaluation 2: Physiological Consistency of Estimated Cognitive Parameters}

Although the parameters obtained through inverse inference reproduce the observed trajectories, it remains unclear whether they track physiological fluctuations in the driver's underlying perceptual and decision-making processes.
Prior research suggests that changes in visual behavior (e.g., increased gaze dispersion~\cite{credidio2012statistical, goodridge2025assessing}, more frequent saccades~\cite{galley1989saccadic}, or shortened fixation duration~\cite{velichkovsky2003visual}) can reflect elevated cognitive strain, uncertainty, and hazard appraisal.
If our inferred parameters capture latent cognitive-state changes (including perceptual uncertainty and looming-averse risk sensitivity), then time periods in which these parameters are flagged as abnormal should co-occur with periods of abnormal visual behavior.
In this section, we extract abnormal segments from both the inferred cognitive-parameter traces and the eye-tracking signals, and quantify their temporal correspondence via overlap-based matching.

\subsubsection{Extracting Abnormal Cognitive Segments.}\label{sec:abnormal_cognitive_segments}
In this analysis, abnormal segments are identified in the time series of three cognitive parameters: $\sigma_{0}$, $\sigma_{\max}$, and $c$.
A statistical thresholding step is applied to flag abnormal frames, followed by signal processing and segment-merging procedures to improve the temporal continuity of the detected segments.

\textbf{Frame-Level Anomaly Detection.}
A frame is defined as cognitively abnormal when the value of a given cognitive parameter exceeds the 90th percentile of that parameter’s within-session distribution, yielding a binary abnormality label sequence (1 = abnormal, 0 = normal) over time.
This percentile-based threshold captures pronounced positive deviations relative to the driver’s baseline within the same session.

\textbf{Signal Processing.}
The binary abnormality label sequence is smoothed by convolution with a Hanning window~\cite{Harris78} (window length $N=20$).

The Hanning window function is defined as:
$$
w(n) = 0.5 \left(1 - \cos\left(\frac{2\pi n}{N-1}\right)\right), \quad n = 0, \dots, N-1,
$$
where $N=20$ in this study.
This smoothing step suppresses isolated frame-level spikes and bridges brief gaps in the abnormality signal.

\textbf{Segment Merging.}
Abnormal cognitive segments are defined as contiguous peaks in the smoothed signal that exceed an amplitude threshold of 0.5. This peak-based merging converts the smoothed frame-level labels into cohesive temporal segments.

\subsubsection{Extracting Abnormal Physiological Segments.}
To enable a direct comparison with the model-derived abnormal parameter segments, abnormal physiological segments are extracted from eye-tracking signals. This section summarizes the selected metrics and the automated rules used to convert frame-level eye-tracking measurements into cohesive abnormal segments.

\textbf{Abnormal Physiological Segments for Perceptual Uncertainty.} 
Perceptual uncertainty is operationalized using gaze transition entropy and fixation stability metrics derived from eye tracking.

\begin{enumerate}
  \item Gaze Entropy
\end{enumerate}

Gaze entropy is quantified using Temporal Gaze Entropy ($H_t$)~\cite{krejtz2015gaze} and Spatial Gaze Entropy ($H_s$)~\cite{goodridge2025assessing}, which capture transition randomness and spatial dispersion, respectively.

To compute these indices, continuous gaze coordinates (X/Y) are discretized into Areas of Interest (AOIs) by partitioning the driver’s field of view into a $3 \times 3$ grid ($N=9$), where each AOI corresponds to a state $i$.

$H_s$ captures how uniformly gaze samples are distributed across AOIs: a higher $H_s$ indicates a more even allocation of attention across regions. It is computed as the Shannon entropy of the AOI occupancy distribution:
\[
H_s(x) = -\sum_{i=1}^{N} p(i) \log_2 p(i),
\]
where $N$ denotes the total number of AOIs, $i$ is the specific AOI index (ranging from 1 to $N$), and $p(i)$ is the proportion (relative frequency) of gaze points falling within AOI $i$.

The value of $H_s$ is normalized by dividing it by the maximum entropy $H_{max} = \log_2 N$, yielding values in $[0,1]$, where 1 corresponds to a uniform distribution across AOIs.

$H_t$ measures the unpredictability of gaze transitions between AOIs under a first-order Markov assumption. A higher $H_t$ indicates less systematic scanning and more random transitions. It is computed as the conditional entropy of the transition probabilities:

\[
H_t(x) = -\sum_{i=1}^{N} p(i) \left[ \sum_{j=1}^{N} p(j|i) \log_2 p(j|i) \right],
\]
where $N$ is the total number of AOIs, $p(i)$ is the stationary-distribution probability (i.e., the total probability of gaze points in state $i$), and $p(j|i)$ is the transition probability from state $i$ to state $j$.

Similar to $H_s$, $H_t$ is normalized by dividing it by $H_{max} = \log_2 N$.

Extremely high entropy values indicate overly scattered, disorganized search that can reduce information efficiency and disrupt sustained visual focus required by time-critical judgments (e.g., time-to-collision)~\cite{credidio2012statistical, goodridge2025assessing}. Extremely low entropy values reflect overly concentrated, tunnel-vision-like viewing, which can reduce awareness of peripheral context and impair situational awareness~\cite{hu2023cognitive}.
Gaze-dispersion abnormalities are then flagged at the frame level using within-session empirical thresholds: frames with entropy values below the 10th percentile or above the 75th percentile are labeled as dispersion-abnormal, and the resulting set is denoted by $A_{\text{disp}}$.

\begin{enumerate}[resume]
  \item Fixation
\end{enumerate}

Fixation stability is quantified from the persistence of near-stationary gaze. Let $(x_t, y_t)$ denote the gaze position at frame $t$. The Euclidean displacement between consecutive samples is:

\[
d_t=\sqrt{(x_{t}-x_{t-1})^2 + (y_{t}-y_{t-1})^2 }.
\]

Fixation frames are encoded by a binary indicator $I_{\text{fix}}[t]$ (1 if $d_t<d_{\text{thr}}$, 0 otherwise), where $d_{\text{thr}}=10$ pixels in our implementation. 

Following~\cite{velichkovsky2003visual}, short or fragmented fixations are treated as reduced visual stability.
Fixation abnormalities are then identified from $I_{\text{fix}}[t]$ using two complementary rules. First, contiguous fixation runs shorter than $\text{thr}_{\text{fix}}=0.1$ s are flagged as short-fixation. Second, the windowed fixation ratio is computed over $W_f=20$ frames as $\text{FixRatio}[t]=\frac{1}{W_f}\sum_{\tau=t-W_f+1}^{t} I_{\text{fix}}[\tau]$ and compared against a within-session threshold $\mu_{\text{ratio}}-\sigma_{\text{ratio}}$ to flag low-fixation-ratio windows, where $\mu_{\text{ratio}}$ and $\sigma_{\text{ratio}}$ are the within-session mean and standard deviation of $\text{FixRatio}[t]$.

A frame is marked as fixation-abnormal if either rule is triggered, and the resulting set is denoted by $A_{\text{fix}}$.

\begin{enumerate}[resume]
  \item Abnormal Physiological Segments
\end{enumerate}

A frame is labeled as perception-abnormal (perceptual uncertainty) if either gaze dispersion or fixation instability is abnormal. The resulting set is:
\[  
A_{\text{physio}} = A_{\text{disp}} \cup A_{\text{fix}}.
\]

The resulting frame-level abnormality sequence is processed using the same Signal Processing (Hanning-window smoothing) and Segment Merging procedures described in Sec.~\ref{sec:abnormal_cognitive_segments}, yielding Abnormal Physiological Segments for Perceptual Uncertainty.

\textbf{Abnormal Physiological Segments for Looming Aversion.} 
Looming aversion is operationalized using saccadic dynamics and pupil dilation change metrics derived from eye tracking.

\begin{enumerate}
  \item Saccadic Dynamics
\end{enumerate}

Saccadic dynamics are quantified using gaze speed derived from inter-frame displacement. Elevated saccadic peak velocity has been associated with increased physiological arousal and urgency~\cite{galley1989saccadic}. 

Let $d_t$ denote the Euclidean displacement between consecutive gaze samples and $f_s$ the sampling frequency. The instantaneous gaze speed is computed as:

\begin{equation*}
    v_t = d_t \cdot f_s.
\end{equation*}

Frames with $v_t$ exceeding the 90th percentile of its within-session distribution are labeled as abnormal-saccade, forming the set $A_{\text{sacc}}$.

\begin{enumerate}[resume]
  \item Pupil Dilation Change
\end{enumerate}

Pupil dilation change is quantified using the temporal derivative of pupil diameter. Rapid pupillary responses to looming stimuli have been reported~\cite{chen2016subliminal}, motivating pupil dynamics as an indicator of looming-related arousal.

Let $\text{rate}[t]$ denote the first-order temporal derivative of pupil diameter. Rapid changes are quantified by Z-score normalization within session:
\begin{equation*}
z_{\text{rate}}[t] = \frac{\text{rate}[t] - \mu_{\text{rate}}}{\sigma_{\text{rate}}},
\end{equation*}
where $\mu_{\text{rate}}$ and $\sigma_{\text{rate}}$ denote the mean and standard deviation of $\text{rate}[t]$ within the session.

Frames exhibiting rapid pupil change are flagged when the standardized rate exceeds a threshold:
\begin{equation*}
\text{fast-change}[t] =
  \begin{cases}
  1, & |z_{\text{rate}}[t]| > z_{\text{thr}}, \\
  0, & \text{otherwise}
  \end{cases},
\end{equation*}
where $z_{\text{thr}} = 1.5$ in our implementation. Frames with $\text{fast-change}[t]\allowbreak = 1$ form the abnormal-pupil set $A_{\text{pupil}}$.

\begin{enumerate}[resume]
  \item Abnormal Physiological Segments
\end{enumerate}

A frame is labeled as looming-abnormal (looming aversion) if either abnormal saccadic dynamics or abnormal pupil dilation change is detected. The resulting set is:
\[
A_{\text{looming}} = A_{\text{sacc}} \cup A_{\text{pupil}}.
\]

The resulting frame-level abnormality sequence is processed using the same Signal Processing (Hanning-window smoothing) and Segment Merging procedures described in Sec.~\ref{sec:abnormal_cognitive_segments}, yielding Abnormal Physiological Segments for Looming Aversion.

\subsubsection{Overlap-Based Alignment Metrics}
Temporal correspondence between cognitively abnormal segments and physiologically abnormal segments is quantified using an automated overlap-based matching procedure. A cognitive segment is counted as matched if it has any temporal overlap with at least one physiological segment. The overall correspondence is summarized by two metrics.

The \textit{match rate} measures the proportion of cognitively abnormal segments that overlap with at least one physiological abnormal segment:
\[
\textit{match rate} = \frac{N_{\text{cog}}^{\text{matched}}}{N_{\text{cog}}^{\text{total}}},
\]
where $N_{\text{cog}}^{\text{matched}}$ is the number of cognitively abnormal segments that overlap with at least one physiological abnormal segment, and $N_{\text{cog}}^{\text{total}}$ is the total number of cognitively abnormal segments.

The \textit{miss rate} measures the proportion of physiological abnormal segments that have no overlap with any cognitively abnormal segment:
\[
\textit{miss rate} = \frac{N_{\text{phys}}^{\text{unmatched}}}{N_{\text{phys}}^{\text{total}}},
\]
where $N_{\text{phys}}^{\text{unmatched}}$ denotes the number of physiologically abnormal segments without any overlapping cognitively abnormal segment, and $N_{\text{phys}}^{\text{total}}$ is the total number of physiologically abnormal segments.

Together, these two metrics provide a concise summary of temporal correspondence between the inferred cognitive-segment abnormalities and the physiological-segment abnormalities.

\subsubsection{Experimental Results} 
Using the overlap-based matching metrics defined above, temporal correspondence between cognitively abnormal segments and physiologically abnormal segments is summarized for each cognitive mechanism. The following results report match/miss rates for perceptual uncertainty and looming aversion.

\textbf{Results of Perceptual Uncertainty.}
Match/miss rates are reported for baseline perceptual noise $\sigma_{0}$ and maximum noise expansion $\sigma_{\max}$ based on overlap between parameter-derived abnormal segments and the abnormal physiological segments for perceptual uncertainty. Results are summarized in two parts: overall alignment performance and robustness across scenario settings.

\begin{enumerate}
  \item Overall Alignment Performance
\end{enumerate}

\begin{figure}[h]
  \centering
  \includegraphics[width=\linewidth]{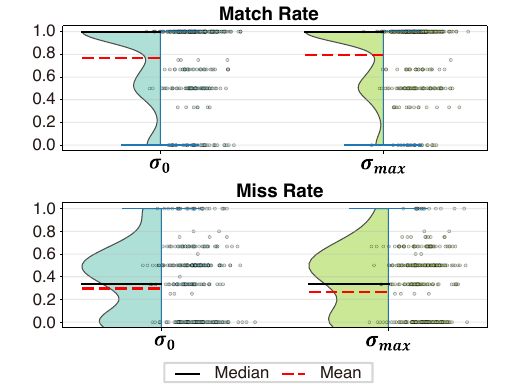}
  \caption{
    Alignment performance between parameter-derived abnormal segments ($\sigma_{0}$ and $\sigma_{\max}$) and abnormal physiological segments for perceptual uncertainty. 
    The upper subplot reports the \textit{match rate}, while the lower subplot reports the \textit{miss rate}.  
    In each panel, the left half shows the distribution as a half-violin plot, and the right half shows the corresponding sample-level scatter; mean and median are overlaid as summary markers.
    \textit{Main take-away:} match rates are typically high, whereas miss rates remain non-negligible, indicating that abnormal segments derived from $\sigma_{0}$ and $\sigma_{\max}$ overlap with many but not all physiological abnormality segments.
}
  \Description{Two stacked plots summarizing alignment between abnormal segments of perceptual-uncertainty parameters (sigma_0, sigma_max) and abnormal eye-tracking segments. The upper plot shows match-rate distribution and the lower plot shows miss-rate distribution; each uses a half-violin distribution on the left and sample-level scatter on the right, with mean/median markers overlaid. Main result: match rates are typically high (mostly around 0.8–1.0; mean > 0.75 and median = 1.0), while miss rates remain non-negligible (often around the mid-range, ~0.5), indicating partial but not complete overlap with physiological abnormality segments.}

  \label{fig:distribution_violin}
\end{figure}

Figure \ref{fig:distribution_violin} summarizes the match/miss rate distributions for $\sigma_{0}$ and $\sigma_{\max}$.

The results showed that the match rates of both cognitive parameters were predominantly distributed in a high range (approximately 0.8–1.0), with mean values above 0.75 and median values equal to 1. 
This indicates that, in most cases, when the model infers a noticeable increase in either $\sigma_{0}$ or $\sigma_{\max}$, reflecting elevated perceptual uncertainty, corresponding abnormalities are simultaneously observed in eye-movement physiology. 
In other words, the internal representation of heightened perceptual uncertainty aligns closely with the eye-movement signature of degraded perceptual processing, including abnormally high or abnormally low gaze entropy and rapid fixation transitions.

Meanwhile, miss rates are concentrated around the mid-range (approximately 0.5). 
This pattern indicates that a substantial portion of abnormal physiological segments do not overlap with parameter-derived abnormal segments. 
These results suggest that $\sigma_{0}$ and $\sigma_{\max}$ capture only part of the physiological manifestation of perceptual uncertainty, with remaining abnormalities potentially attributable to additional mechanisms not explicitly represented in the current model.

In summary, $\sigma_{0}$ and $\sigma_{\max}$ demonstrate consistent temporal correspondence with abnormal eye-movement segments linked to perceptual uncertainty, supporting their use as indicators of this mechanism.

\begin{enumerate}[resume]
  \item Alignment Robustness Across Scenario Settings
\end{enumerate}

\begin{figure}[h]
  \centering
  \includegraphics[width=\linewidth]{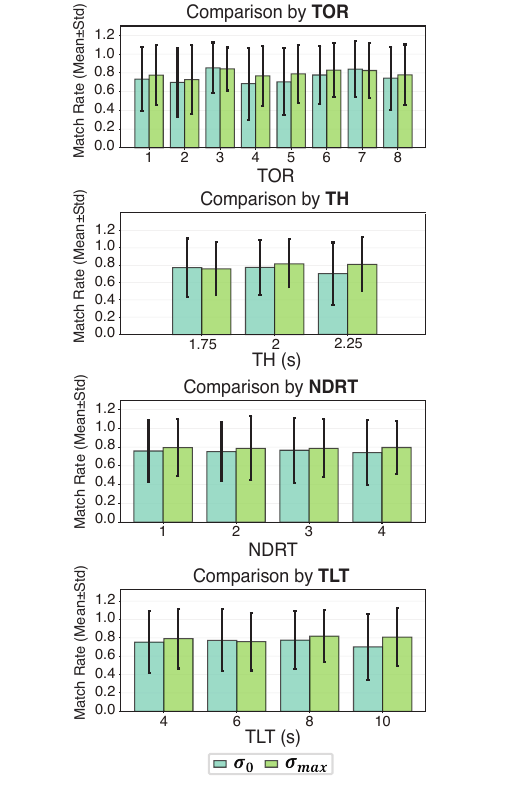}
  \caption{
    Comparison of the match rate across different scenario settings. 
    Bars report the mean match rate for each condition, with error bars indicating the standard deviation, 
    for two cognitive parameters --- baseline perceptual noise ($\sigma_{0}$) 
    and maximum noise expansion ($\sigma_{\max}$). 
    The results are categorized based on the following factors:
    \textbf{Takeover Request (TOR)}, with configurations defined in Table \ref{tab:tor_configurations}.
    \textbf{Time Headway (TH)}, with three settings: 1.75~s, 2~s, and 2.25~s.
    \textbf{Non-Driving Task (NDRT)}, categorized into four conditions: 
      1-No NDRT, 
      2-Watching a tablet video, 
      3-Playing the smartphone game "2048", 
      4-Reading a book.
    \textbf{Takeover Lead Time (TLT)}, with four time intervals: 4~s, 6~s, 8~s, and 10~s.
    \textit{Main take-away:} match rates for $\sigma_{0}$ and $\sigma_{\max}$ are broadly stable across TH, NDRT, and TLT, while TOR shows more visible between-condition differences.
}
  \Description{A grouped bar chart comparing mean match rate (with standard-deviation error bars) across pre-takeover scenario settings for two parameters, sigma_0 and sigma_max. Categories include TOR configurations, time headway (TH), non-driving task (NDRT) types, and takeover lead time (TLT); the figure highlights relatively stable match rates across TH/NDRT/TLT and more visible differences across TOR.}
  \label{fig:comparison}
\end{figure}

To evaluate the robustness of the cognitive--physiological alignment under different initial experimental conditions (as defined in Study~1, including variations in TOR, TH, NDRT, and TLT), match rates were compared across these scenario categories. Figure \ref{fig:comparison} summarizes the resulting match rates by condition. 

Across TH, NDRT, and TLT settings, both $\sigma_{0}$ and $\sigma_{\max}$ show consistently high and similar match rates, with limited variation across levels. In contrast, match rates differ more visibly across TOR conditions, suggesting that alternative TOR strategies may shift the strength of parameter--physiology alignment.

To assess whether differences in match rate across initial configuration conditions were practically small, a margin-based comparison was conducted with an equivalence-style tolerance of $\Delta = 0.10$. Specifically, the maximum between-condition mean difference (MaxDiff) was compared against $\Delta$, and one-way ANOVA $p$-values are reported as a complementary check for detectable differences (Table~\ref{tab:equivalence}). 

For TH, NDRT, and TLT, MaxDiff remained within the tolerance (MaxDiff $< 0.08$), suggesting that match-rate differences were small in magnitude for both $\sigma_{0}$ and $\sigma_{\max}$ across these three dimensions; correspondingly, the ANOVA results did not detect differences ($p > 0.54$ for all tests).

In contrast, for TOR, MaxDiff exceeded $\Delta$ for both parameters (MaxDiff $\ge 0.11$), so the match-rate differences were not uniformly bounded within the same tolerance, even though the ANOVA tests remained non-significant ($p > 0.13$).

\begin{table}[h]
  \centering
\caption{ANOVA $p$-values and maximum between-condition mean differences (MaxDiff) for \textit{match rate} across pre-takeover scenario settings.}
% \resizebox{\columnwidth}{!}{%
  \begin{tabular}{ccccc}
  \toprule
  \textbf{Dimension} & \textbf{Parameter} & \textbf{F} & \textbf{p-value} & \textbf{MaxDiff} \\
  \midrule
  \multirow{2}{*}{TH} & $\sigma_{0}$ & 0.71 & 0.5465 & 0.074 \\
   & $\sigma_{\max}$ & 0.65 & 0.5859 & 0.059 \\
  \midrule
  \multirow{2}{*}{NDRT} & $\sigma_{0}$ & 0.09 & 0.9670 & 0.025 \\
   & $\sigma_{\max}$ & 0.02 & 0.9966 & 0.009 \\
  \midrule
  \multirow{2}{*}{TLT} & $\sigma_{0}$ & 0.53 & 0.6384 & 0.064 \\
   & $\sigma_{\max}$ & 0.41 & 0.7524 & 0.036 \\
  \midrule
  \multirow{2}{*}{TOR} & $\sigma_{0}$ & 1.58 & 0.1397 & 0.169 \\
   & $\sigma_{\max}$ & 0.61 & 0.7442 & 0.114 \\
  \bottomrule
  \end{tabular}
  \label{tab:equivalence}
  \end{table}
  
Overall, these results indicate that match-rate differences for perceptual uncertainty were bounded within $\Delta$ across TH, NDRT, and TLT, whereas TOR showed larger between-condition variation (MaxDiff $>\Delta$). This pattern suggests that the parameter--physiology alignment is relatively stable across most scenario dimensions, with TOR potentially associated with greater variability in alignment strength.

\textbf{Results of Looming Aversion.}
Match/miss rates are reported for the looming-aversion weighting coefficient ($c$) based on overlap between parameter-derived abnormal segments and the abnormal physiological segments associated with looming-related responses. Results are summarized in two parts: overall alignment performance and robustness across scenario settings. 

\begin{enumerate}
  \item Overall Alignment Performance
\end{enumerate}

\begin{figure}[h]
  \centering
  \includegraphics[width=\linewidth]{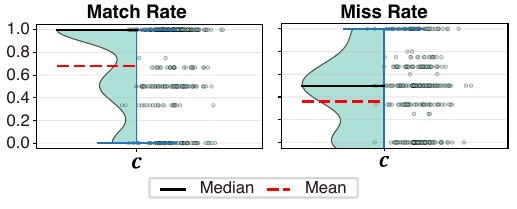}
  \caption{
    Alignment performance between parameter-derived abnormal looming-aversion segments ($c$) and abnormal physiological segments associated with looming-related responses. 
    The left subplot reports the \textit{match rate}, while the right subplot reports the \textit{miss rate}. 
    In each panel, the left half shows the distribution as a half-violin plot, and the right half shows the corresponding sample-level scatter; mean and median are overlaid as summary markers.
    Subplots use the same y-axis scale; tick labels are shown on the left subplot only.
    \textit{Main take-away:} match rates are generally moderate to high, whereas miss rates remain non-negligible, indicating that $c$-derived abnormal segments overlap with many but not all physiological abnormality segments.
}
  \Description{Two side-by-side plots summarizing alignment between abnormal segments of looming-aversion parameter c and abnormal physiological segments. The left plot shows match-rate distribution and the right plot shows miss-rate distribution; each uses a half-violin plus sample scatter with mean/median markers, on a shared y-axis scale. Main result: match rate is generally moderate to high (median = 1.0; mean ~ 0.66), whereas miss rate is still substantial (mean ~ 0.40; median ~ 0.50), suggesting c captures a subset of looming-related physiological abnormalities.}

  \label{fig:c_similarity_distribution_violin}
\end{figure}

Figure \ref{fig:c_similarity_distribution_violin} summarizes the match/miss rate distributions for $c$.

The match rate results indicate that the alignment between $c$-abnormal segments and physiological abnormalities is generally moderate to high, with the median reaching 1.0, while the mean is approximately 0.66. This shows that in the majority of moments where the model infers heightened looming sensitivity—interpreted as increased perceived threat from rapidly approaching objects—the driver's eye-movement signals also exhibit intensified responses associated with tension, such as rapid saccades or strong pupil changes.

However, miss rates remain substantial (mean $\approx 0.40$, median $=0.50$), indicating that many physiologically abnormal segments have no temporal overlap with $c$-derived abnormal segments. This pattern suggests that looming-aversion estimates captured by $c$ correspond to a subset of physiological responses, with remaining abnormalities potentially reflecting additional processes beyond looming-related threat sensitivity.

In summary, the looming-aversion weighting coefficient ($c$) demonstrates consistent temporal correspondence with a subset of physiologically abnormal segments associated with looming-related visual tension, supporting its use as an indicator of this mechanism.

\begin{enumerate}[resume]
  \item Alignment Robustness Across Scenario Settings
\end{enumerate}

\begin{figure}[h]
  \centering
  \includegraphics[width=\linewidth]{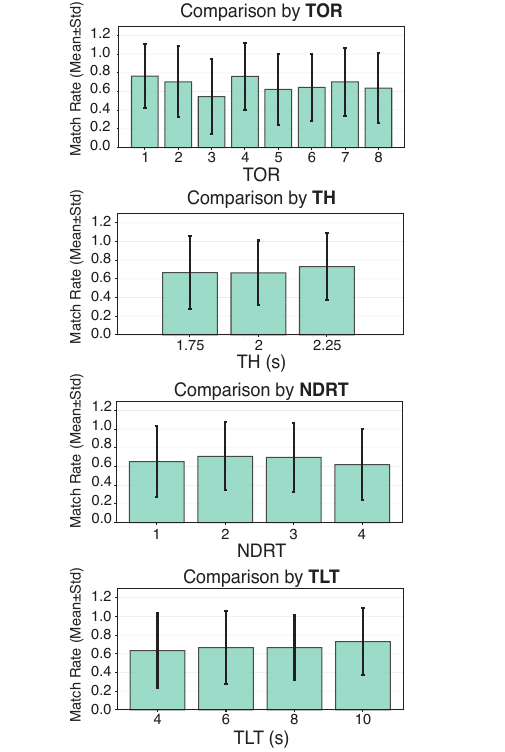}
  \caption{
    Comparison of the match rate across different scenario settings. 
    Bars report the mean match rate for each condition, with error bars indicating the standard deviation, 
    for the looming-aversion weighting coefficient ($c$). 
    The results are categorized based on the following factors:
    \textbf{Takeover Request (TOR)}, with configurations defined in Table \ref{tab:tor_configurations}.
    \textbf{Time Headway (TH)}, with three settings: 1.75~s, 2~s, and 2.25~s.
    \textbf{Non-Driving Task (NDRT)}, categorized into four conditions: 
      1-No NDRT, 
      2-Watching a tablet video, 
      3-Playing the smartphone game "2048", 
      4-Reading a book.
    \textbf{Takeover Lead Time (TLT)}, with four time intervals: 4~s, 6~s, 8~s, and 10~s.
    \textit{Main take-away:} match rates for $c$ are largely consistent across TH, NDRT, and TLT, while TOR exhibits more pronounced between-condition differences.
}
  \Description{A bar chart comparing mean match rate (with standard-deviation error bars) across scenario settings for the looming-aversion weighting coefficient c. Categories include TOR, TH, NDRT, and TLT; the figure emphasizes largely consistent match rates across TH/NDRT/TLT and more pronounced between-condition differences across TOR.}
  \label{fig:c_comparison}
\end{figure}

To evaluate the robustness of the cognitive--physiological alignment under different initial experimental conditions (as defined in Study~1, including variations in TOR, TH, NDRT, and TLT), match rates were compared across these scenario categories. Figure \ref{fig:c_comparison} summarizes the resulting match rates by condition. 

Across TH, NDRT, and TLT, match rates appear similar, with closely aligned means and overlapping standard-deviation ranges. This pattern suggests limited variation in the $c$-based alignment under these environmental configurations and task demands, whereas TOR shows more pronounced between-condition differences.

To assess whether between-condition differences were practically small, a margin-based comparison was conducted with an equivalence-style tolerance of $\Delta = 0.10$. Specifically, the maximum between-condition mean difference (MaxDiff) was compared against $\Delta$, and one-way ANOVA $p$-values are reported as a complementary check for detectable differences (Table~\ref{tab:equivalence_bias_inverse_tta}). For TH, NDRT, and TLT, MaxDiff remained within the tolerance (MaxDiff $\le 0.096$), suggesting small match-rate differences; correspondingly, the ANOVA results did not detect differences ($p > 0.38$ for all tests). In contrast, for TOR, MaxDiff exceeded $\Delta$ (MaxDiff $= 0.221$), so the differences were not uniformly bounded within the same tolerance, even though the ANOVA test remained non-significant ($p = 0.127$).

Together, these findings demonstrate that the looming-aversion parameter $c$ exhibits strong robustness under different TH, NDRT, and TLT levels. At the same time, variations in TOR wording may subtly influence how behavioral estimates of looming sensitivity align with physiological indicators, suggesting that drivers' expectations regarding takeover instructions may influence their perceived threat responses prior to control transition.

\begin{table}[h]
  \centering
  \caption{ANOVA $p$-values and maximum between-condition mean differences (MaxDiff) for \textit{match rate} across pre-takeover scenario settings.}
  % \resizebox{\columnwidth}{!}{%
  \begin{tabular}{ccccc}
  \toprule
  \textbf{Dimension} & \textbf{Parameter} & \textbf{F} & \textbf{p-value} & \textbf{MaxDiff} \\
  \midrule
  TH & $c$ & 0.76 & 0.5169 & 0.096 \\
  NDRT & $c$ & 1.01 & 0.3876 & 0.089 \\
  TLT & $c$ & 0.86 & 0.7548 & 0.093 \\
  TOR & $c$ & 1.63 & 0.1272 & 0.221 \\
  \bottomrule
  \end{tabular}
  % } 
  \label{tab:equivalence_bias_inverse_tta}
\end{table}

\subsection{Qualitative Evaluation}

To analyze and illustrate how our method simulates driver behavior dynamically and forecasts risk, we randomly selected a takeover episode that culminated in a collision for fine-grained analysis. In this experiment, the TLT was 8 s. Approximately 1 s after hearing the takeover prompt, the driver depressed the brake and assumed control at roughly 190 m upstream of the work zone. The driver then executed avoidance maneuvers over the remaining distance. Our behavioral simulation for this case is confined to this interval, enabling a focused examination of the model’s moment-to-moment predictions and the evolving risk profile during the early stage after takeover.

\subsubsection{Prediction Accuracy}

\begin{figure}[h]
  \centering
  \includegraphics[width=\linewidth]{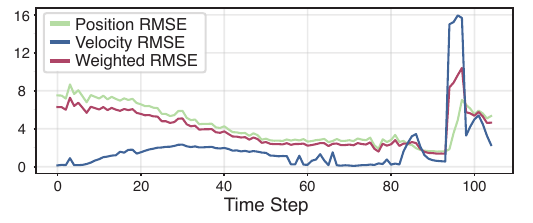}
  \caption{Root mean squared error (RMSE) of position and velocity prediction errors over time step.}
  \Description{A line plot of prediction error over time steps during a takeover episode, reporting RMSE for position and velocity (and/or an aggregated RMSE). Errors generally decrease as the adaptive model updates, with a sharp spike around the collision time due to post-impact divergence.}
  \label{fig:rmse_results}
\end{figure}

At each rolling step, the model predicts the next $H=30$ steps (3 s). This yields paired sequences of simulated human-like behavior $\{\hat{\mathbf{p}}_{k+h}, \hat{\mathbf{v}}_{k+h}\}_{h=1}^{H}$ and the driver's realized trajectory $\{\mathbf{p}_{k+h}, \mathbf{v}_{k+h}\}_{h=1}^{H}$, from which we compute dynamic RMSE over the horizon:
\begin{equation*}
    \mathrm{RMSE}_{\text{pos}}(k) = \sqrt{\frac{1}{H}\sum_{h=1}^{H}\left\lVert \hat{\mathbf{p}}_{k+h}-\mathbf{p}_{k+h}\right\rVert_2^{2}},
\end{equation*}
\begin{equation*}
    \mathrm{RMSE}_{\text{vel}}(k) = \sqrt{\frac{1}{H}\sum_{h=1}^{H}\left\lVert \hat{\mathbf{v}}_{k+h}-\mathbf{v}_{k+h}\right\rVert_2^{2}}.
\end{equation*}

To summarize both spatial and kinematic accuracy, we report a weighted RMSE:
\begin{equation*}
  \begin{split}
    \mathrm{RMSE}_{w}(k)
    &= \sqrt{\alpha\,\mathrm{RMSE}_{\text{pos}}(k)^{2}
    + \beta\,\mathrm{RMSE}_{\text{vel}}(k)^{2}},\\
    &\qquad \alpha,\beta\ge 0,\ \alpha+\beta=1,
  \end{split}
\end{equation*}
where $\alpha$ and $\beta$ are weights reflecting the relative importance of position and velocity errors.

Figure \ref{fig:rmse_results} illustrates the RMSE of position and velocity prediction errors over time steps. As time progresses, with adaptive cognitive parameters, the 3 s--ahead prediction error steadily declines: by T=40 (4 s), the position RMSE stabilizes within 3 m and the velocity RMSE remains below 1 m/s. A sharp spike appears around T=90 (9 s), coinciding with the collision; post-impact dynamics (large impulses and rebound) differ between the real experiment and the simulator, leading to pronounced divergence immediately after impact. Overall, these results indicate that online adaptation of cognitive parameters improves short-horizon predictive accuracy during the takeover episode, corroborating the role of dynamic inference in tracking rapidly changing driver behavior.

\begin{figure*}[h]
  \centering
  \includegraphics[width=0.95\linewidth]{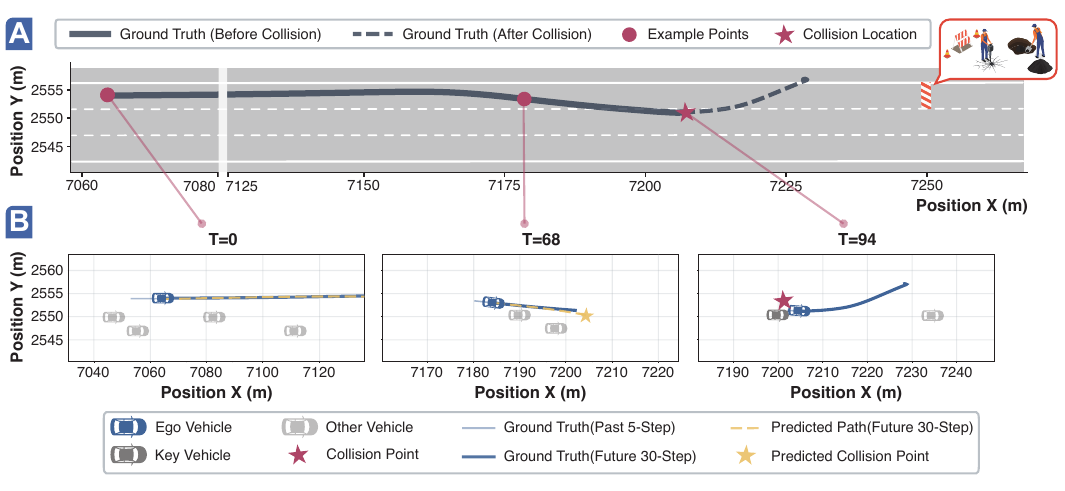}
  \caption{Cognition-aware prediction model evaluation during a takeover episode. (A) Spatial trajectory showing the driver takeover position, lane change maneuver, and collision location. (B) Three temporal snapshots illustrating the evolution of predicted risk and behavioral dynamics: takeover initiation (T=0), early collision warning (T=68), and actual collision (T=94).}
  \Description{Cognition-aware prediction evaluation for one takeover episode. Panel (A) is a spatial trajectory map marking the takeover position, the lane-change maneuver path, and the collision location. Panel (B) shows three temporal snapshots (T=0 takeover initiation, T=68 early collision warning, T=94 collision) illustrating the evolution of predicted risk and behavioral dynamics.}
  \label{fig:results}
\end{figure*}

\subsubsection{Timeline of Early Collision Anticipation}

We now analyze this takeover process in greater detail, focusing on the temporal evolution of behavior and risk.

Figure \ref{fig:results}A contextualizes the episode within which the driver takes over at $X=7060\,\mathrm{m}$ ($T=0$), with a work zone ahead in the same lane at $X=7250\,\mathrm{m}$. After takeover, the ego vehicle continues straight for a short distance and initiates a lane change at $X=7160\,\mathrm{m}$ toward the adjacent (right) lane. A collision with a rear-approaching vehicle in that adjacent lane occurs at $X=7205\,\mathrm{m}$ ($T=94$), i.e., $9.4\,\mathrm{s}$ after takeover. 

To explain the evolving risk, Figure \ref{fig:results}B samples three snapshots: the takeover moment ($T=0$), the first risk flag ($T=68$), and the collision ($T=94$). At $T=0$, adjacent-lane traffic is relatively dense and the driver is likely still in a transitional response phase; the ego maintains a straight path over the next $3\,\mathrm{s}$, which the model predicts accurately. At $T=68$, the model begins forecasting a potential collision within the next $3\,\mathrm{s}$ near $X=7205\,\mathrm{m}$ with the rear-approaching vehicle in the adjacent lane---26 simulation steps earlier than the impact at $T=94$ (a $2.6\,\mathrm{s}$ lead), and beyond what a constant-velocity heuristic could infer given the ego's instantaneous speed and heading. Notably, this alert arrives $0.8\,\mathrm{s}$ earlier than a vanilla PPO baseline without cognitive parameters. Subsequent rollouts continue to mark elevated risk at the eventual impact location. 

Taken together, this case indicates that integrating adaptive cognitive parameters extends the risk-detection horizon during takeover, enabling earlier collision warnings and supporting timely driver alerts and, when necessary, escalation to lower-level safety safeguards in human–vehicle interaction.

\subsubsection{Cognitive–Behavior Coupling}

\begin{figure*}[h]
  \centering
  \includegraphics[width=0.95\linewidth]{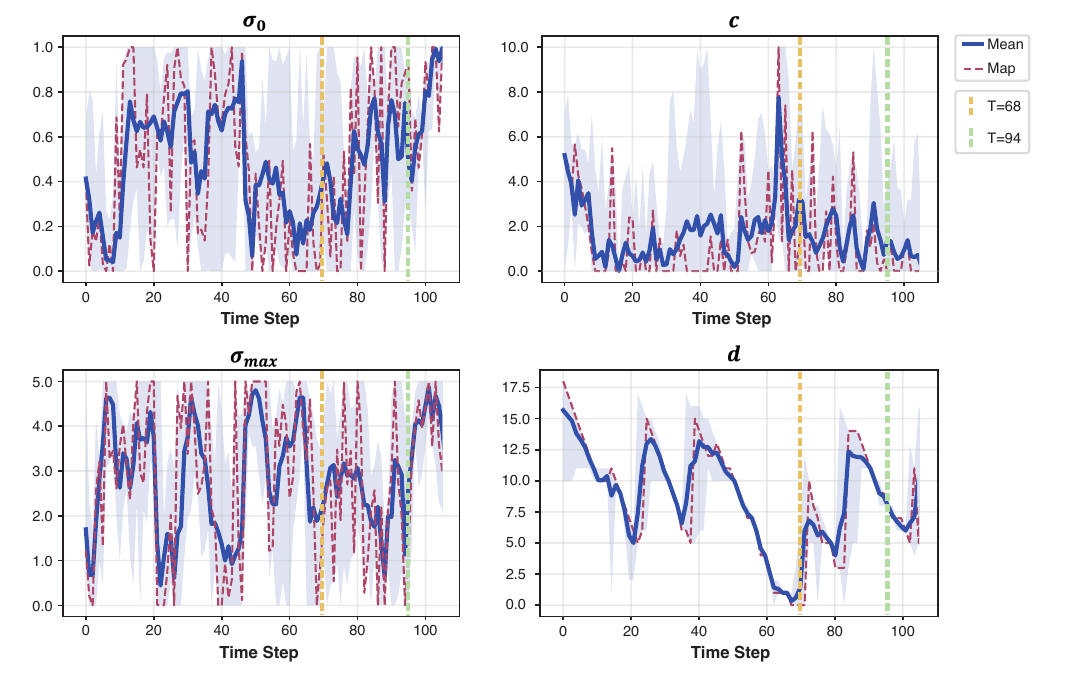}
  \caption{Temporal evolution of inferred cognitive parameters during the takeover episode. The cognitive parameters are estimated using a 5-step sliding window, showing the dynamic adaptation of looming-aversion weight, action delay, near-range noise floor, and far-range noise scaling throughout the collision scenario.}
  \Description{Time-series plots of inferred cognitive parameters across the takeover episode, estimated using a 5-step sliding window. Curves show dynamic adaptation of looming-aversion weight c, action delay d, near-range perceptual noise floor sigma_0, and far-range noise scaling/saturation sigma_max over time.}
  \label{fig:results_cog}
\end{figure*}

\begin{figure}[h]
  \centering
  \includegraphics[width=\linewidth]{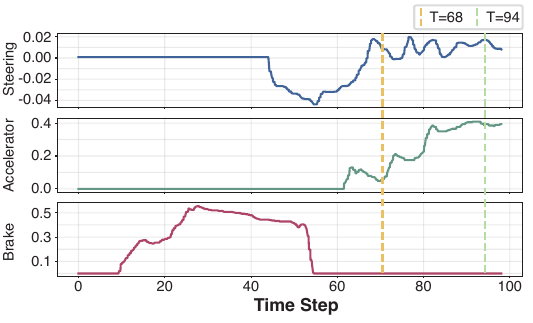}
  \caption{Driver control inputs throughout the takeover episode. The time series shows steering angle, accelerator pedal position, and brake pedal position, revealing the temporal dynamics of human control actions during the collision scenario.}
  \Description{Time-series of driver control inputs during the takeover episode, including steering angle, accelerator pedal position, and brake pedal position, showing how longitudinal and lateral control actions change over time in the collision scenario.}
  \label{fig:results_action}
\end{figure}

Building on the previous sections where we demonstrated that cognitive adaptation improves dynamic regulation of driver behavior and enables effective earlier risk warnings, we further examine how the driver's actions couple with the inferred cognitive state over time. 

Figure \ref{fig:results_cog} plots the online estimates of four cognitive-state parameters obtained using a past-5-step window, while Figure \ref{fig:results_action} shows the driver's steering, accelerator, and brake inputs throughout the episode. 

At the moment of takeover when control transitions from vehicle to human, both the looming-aversion weight $c$ and the action delay $d$ are at relatively high values. A higher $c$ is consistent with the driver's attention being anchored to the same-lane work zone rapidly approaching in the forward field of view; a higher $d$ reflects post-disengagement sluggishness from the NDRT, matching the one-second (10 steps) period with no control input visible in Figure \ref{fig:results_action}. The first deliberate action is a brake application, further corroborating the interpretation that the driver's early attention is dominated by the looming hazard ahead.

As the driver settles, the near-range noise floor $\sigma_0$ and both the looming-aversion weight $c$ and action delay $d$ decrease, indicating improved self-state and growing command of the surrounding context following real-time deceleration and a brief buffer of several seconds. The driver then initiates a rightward lane change.

Around the moment when the model first forecasts the collision, the inferred cognitive state shifts. The near-range noise floor $\sigma_0$ begins a sustained rise, signaling elevated baseline perceptual noise and more volatile risk appraisal; consequently, the driver's assessment of the far-rear threat in the target lane becomes less reliable. In parallel, the continued increase of $d$ suggests lengthening perception-to-action latency. In this condition, the driver does not accurately register the rearward hazard and, after committing to the lane change, accelerates sluggishly: at $T=68$, even after the nearby vehicle in the adjacent lane has moved away, there is no decisive throttle application, only a brief, light press that is insufficient to avoid the impending collision with the rear-approaching vehicle.

Synthesizing behavior with the online-inferred cognitive state, we find that the estimated states co-vary with control actions in a way that both explains the observed behavior and supports early risk warnings. In particular, the sustained rise of $\sigma_0$ and $d$ from $T=68$ onward signals a shift toward noisier perception and longer perception–action latency. Taken together with the prediction layer's collision flag and the cognitive layer's indication of perceptual degradation and delay, these cues suggest an impending loss of safety margin. On this basis, the interaction system can provide timely, cognition-aware assistance targeting perception and delay, before risk fully materializes.

\subsubsection{Physiological Reflection of Cognitive Parameters}

\begin{figure*}[t]
    \centering
    \includegraphics[width=1\textwidth]{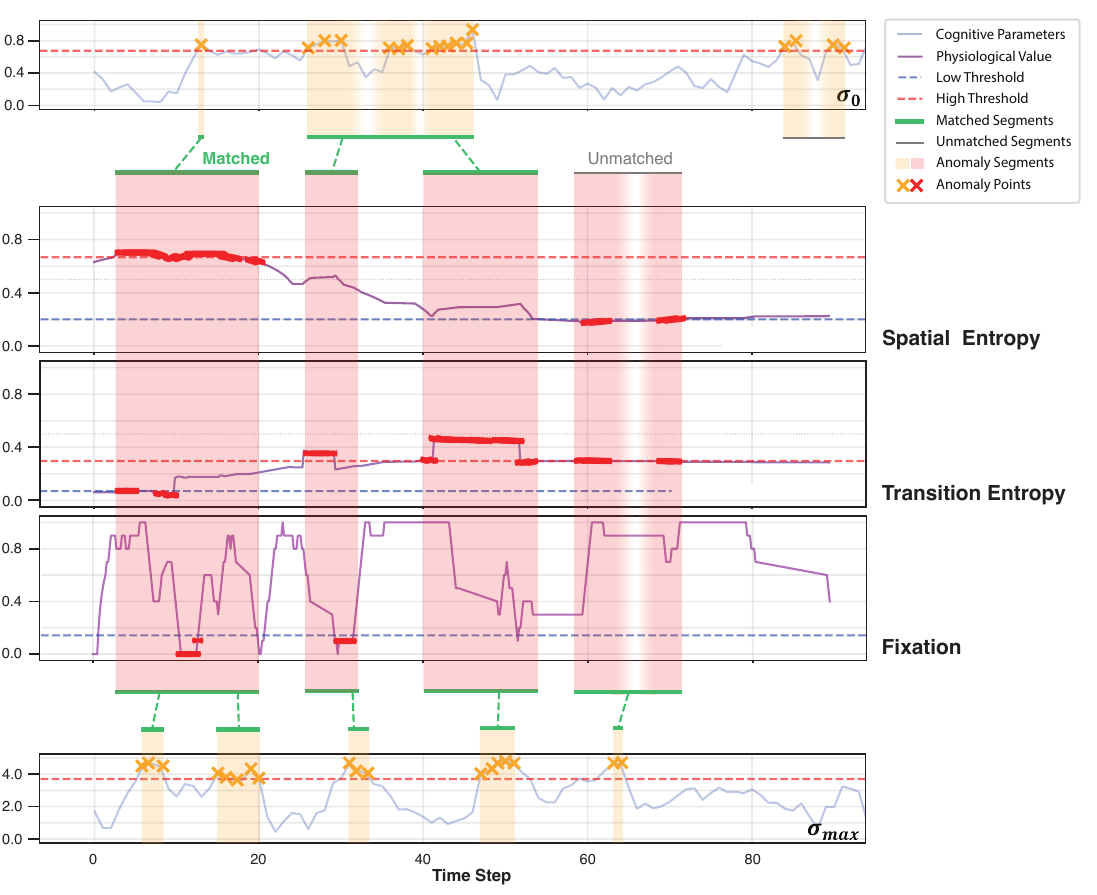}
    \caption{This figure presents the relationship between abnormal segments of the perceptual noise parameters, $\sigma_0$ and $\sigma_{\max}$, and the corresponding physiological data segments, including spatial entropy, transition entropy and fixation. The top plot shows the time series of the $\sigma_0$ parameter, and the bottom plot represents the time series of the $\sigma_{\max}$ parameter. The middle three plots display the time series of the physiological data, with red shading marking abnormal segments in the physiological data, and yellow shading indicating abnormal segments in the cognitive parameters. The gradient regions represent the ends of anomalies identified as part of the same segment due to Hanning window processing. Green horizontal lines indicate segments where the cognitive and physiological data align (matched segments), while gray lines represent segments where no alignment is found (unmatched segments).}
    \Description{A multi-panel time-series figure relating abnormal segments in perceptual-noise parameters sigma_0 (top) and sigma_max (bottom) to abnormal segments in physiological metrics (middle panels: spatial entropy, transition entropy, and fixation). Red shading marks abnormal physiological segments, yellow shading marks abnormal cognitive-parameter segments, gradient regions indicate anomaly boundaries after Hanning-window processing, and horizontal markers distinguish matched (green) versus unmatched (gray) segments.}
    \label{fig:match_sigma}
\end{figure*}

\begin{figure*}[t]
    \centering
    \includegraphics[width=1\textwidth]{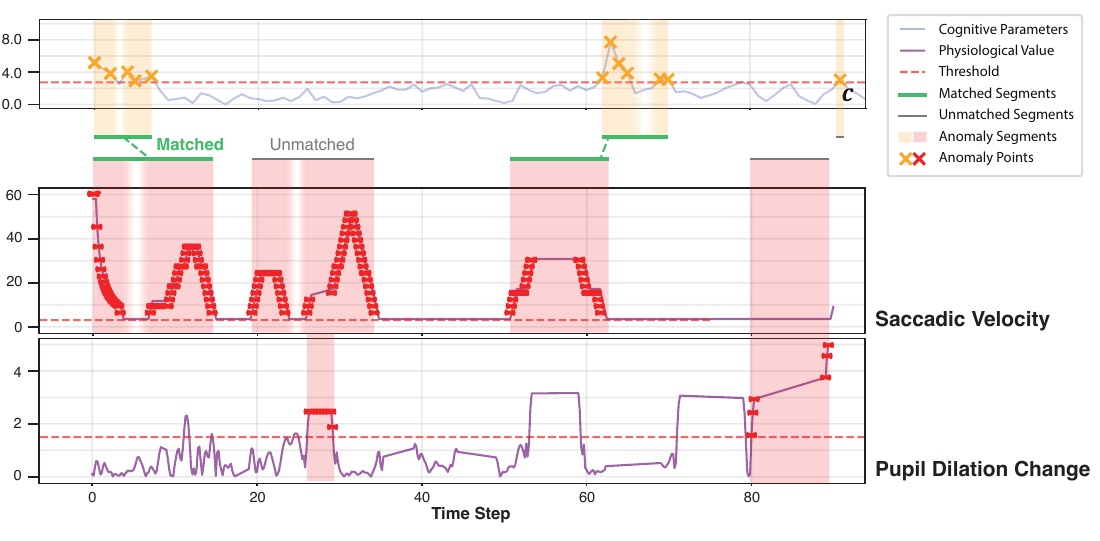}
    \caption{This figure presents the relationship between abnormal segments of the looming aversion weighting coefficient, $c$, and the corresponding physiological data segments, including saccadic velocity and pupil dilation changes. The top plot shows the time series of the $c$ parameter, and the other two plot represents the time series of the physiological data, with red shading marking abnormal segments in the physiological data, and yellow shading indicating abnormal segments in the cognitive parameters. The gradient regions represent the ends of anomalies identified as part of the same segment due to Hanning window processing. Green horizontal lines indicate segments where the cognitive and physiological data align (matched segments), while gray lines represent segments where no alignment is found (unmatched segments).}
    \Description{A multi-panel time-series figure relating abnormal segments in looming-aversion parameter c (top) to abnormal segments in physiological signals (lower panels: saccadic velocity and pupil-dilation changes). Red shading marks abnormal physiological segments, yellow shading marks abnormal cognitive-parameter segments, gradient regions indicate anomaly boundaries after Hanning-window processing, and horizontal markers distinguish matched (green) versus unmatched (gray) segments.}
    \label{fig:match_c}
\end{figure*}

Figure~\ref{fig:match_sigma} and Figure~\ref{fig:match_c} illustrate the temporal correspondence between inferred cognitive parameters and physiological indicators during a representative takeover episode. Overall, fluctuations in perceptual uncertainty and looming aversion exhibit clear alignment with abnormal patterns in eye-tracking and pupillometry, providing supporting evidence that the inferred parameters capture cognition-related changes rather than merely fitting observed trajectories.

As shown in Figure~\ref{fig:match_sigma}, $\sigma_{\max}$ demonstrates stronger and more consistent alignment with physiological abnormalities than $\sigma_0$. While $\sigma_0$ exhibits three abnormal segments, only two overlap with physiological events, and one physiological abnormal interval remains unmatched. In contrast, $\sigma_{\max}$ produces five abnormal segments, all of which coincide with physiological abnormalities, suggesting that far-range perceptual uncertainty is more sensitive to transient cognitive disruptions in this episode.

Figure~\ref{fig:match_c} shows the correspondence between looming aversion $c$ and saccadic velocity and pupil dilation. The first two abnormal segments of $c$ align with major physiological deviations, indicating that changes in looming sensitivity co-occur with heightened visual and arousal responses. Although not every physiological anomaly is mirrored by $c$, the dominant segments are captured, consistent with its role in modulating risk-related decision-making.

Taken together, these results suggest that the inferred cognitive parameters exhibit meaningful temporal coupling with independent physiological measures. This alignment supports the interpretability of the proposed cognitive modeling framework and indicates that online parameter inference reflects underlying cognitive dynamics rather than solely compensating for trajectory prediction errors.

\section{Discussion}

\subsection{Interpretability of Cognitive Parameters on Behavior}
In this single-episode analysis, the trace of $\sigma_{\max}$ varies over time. Given that the geometry is dominated by near-/mid-range interactions with a rear-approaching vehicle in the adjacent lane, and that inference uses a short identification window ($L{=}5$) with a 3 s look-ahead ($H{=}30$), far-range evidence is only weakly excited. In this setting, three parameters carry the main explanatory weight for short-horizon foresight:

(i) \textit{Near-range reliability ($\boldsymbol{\sigma_0}$) and timing ($\boldsymbol{d}$) as foundational states.} $\sigma_0$ governs the credibility of immediate perceptual input that drives the rolling likelihood, while $d$ sets the perception–action latency that shapes when control actually arrives. Together they determine whether braking and throttle adjustments occur early enough within a 2 s horizon to preserve margin. The observed pattern—an initial no-input interval followed by braking, and later sluggish throttle after lane-change commitment—is consistent with elevated $d$ and rising $\sigma_0$.

(ii) \textit{Proximity effects that amplify $\boldsymbol{\sigma_0}$.} Because headways are short, small increases in $\sigma_0$ disproportionately affect risk forecasts: near-field state estimates become less stable, predictive variance grows, and the driver’s acceleration becomes conservative—matching the hesitant re-engagement of throttle even when the adjacent vehicle has moved away.

(iii) \textit{Looming sensitivity ($\boldsymbol{c}$) as a modulator of action selection.} $c$ shapes responsiveness to closing-speed cues. Its high value at takeover explains prompt braking and a cautious posture toward the rapidly approaching work zone.

In short, within a near-/mid-range takeover episode, $\sigma_0$ and $d$ form the backbone of short-horizon prediction by anchoring perceptual reliability and actuation timing, while $c$ modulates attention to looming hazards that tilt the balance between braking and acceleration. This triad explains both the moment-to-moment behavior and the emergence of early collision warnings.

\subsection{Model Stability Analysis}

In the analysis provided in the Quantitative Evaluation 2 section, we compared the performance of the model across different initial experimental settings, focusing on how well the abnormal segments of model parameters aligned with the abnormal segments of physiological data. The results showed that among the four experimental settings—TOR strategies, TH, NDRT, and TLT—TH, NDRT, and TLT did not significantly impact the model's accuracy. In contrast, the model’s performance across different TOR strategies showed larger between-condition variation though not statistically significant. This suggests that TOR strategy might have a distinct effect on model stability, unlike the other three settings.

The reason for this distinction lies in how each setup interacts with the model. TH, NDRT, and TLT directly affect the scenario's state. These factors influence the model indirectly through the observation, as the observation serves as the input to the policy network, thereby shaping the model's decision-making process. While these factors do not directly enter the model as independent variables, their influence on the state transitions via the observation mechanism ensures that the model can adapt to changes in the environment. In contrast, the TOR strategy does not affect the scenario's state or the observation. It influences the driver's internal state, such as cognitive load, attention, or readiness to take over. This influence is more psychological and less environmental, meaning it cannot be processed by the model in the same way the other factors can.

As a result, the TOR strategy introduces variability into the model's accuracy by altering the driver’s state, which ultimately affects the model’s output. This underscores the idea that in models of bounded rationality, certain initial experimental settings can influence the model indirectly through the observation, while factors like the TOR strategy that do not directly impact the observation must be treated as independent variables and used as prior inputs. This distinction is crucial for understanding model stability and ensuring that TOR strategies, which influence the driver’s cognition, are properly incorporated into the model's structure as separate input components.

\subsection{Cognition-Driven Design: Let Causes Shape Assistance}
Our results link what the model foresees to why risk emerges. In the single-episode analysis, rising $\sigma_0$ (near-range perceptual unreliability) and increasing $d$ (perception–action delay) precede the collision and co-occur with non-normative control (an initial no-input interval, prompt braking, then hesitant throttle after lane-change commitment). At the corpus level, earlier warnings and longer lead-time coverage further indicate that cognitive trends become informative before trajectory geometry alone makes the hazard obvious. Taken together, the warning is not merely a signal to "beep"; it is a time window to stage interventions.

This causal understanding fundamentally reshapes how collision warnings should function. Rather than passive alerts, warnings become opportunities for cognition-driven intervention that addresses the root causes of emerging risk. The key insight is that cognitive parameters are not only explanatory; they are control-relevant and map directly to targeted assistance policies that can mitigate specific cognitive limitations as they develop:

% \begin{itemize}
    When near-range reliability ($\boldsymbol{\sigma_0}$) deteriorates—indicating the driver is "seeing less stably, judging less precisely"—the optimal response is to slow the interaction and add redundancy. This involves minimizing nonessential displays and foregrounding the smallest useful set tied to the current task (e.g., rearward vehicle risk), while simultaneously inflating recommended safety margins to preserve reaction time and de-emphasizing aggressive acceleration suggestions alongside raising lane-change confirmation thresholds.

    Similarly, when action delay ($\boldsymbol{d}$) increases, the fundamental challenge shifts from determining what to do to ensuring actions occur in time. This necessitates prioritizing buffering interventions, including brief automation hold, throttle suppression or brake pre-load, and feed-forward cues that announce the next step, all of which serve to bridge the latency in the human loop.

    Finally, elevated looming sensitivity ($\boldsymbol{c}$) reflects heightened sensitivity to closing-speed cues, which at takeover manifests as prompt braking and a cautious posture toward rapidly approaching hazards such as work zones. Under these conditions, the system should provide clearer spatial references and reduce visual clutter to support more accurate distance judgment, thereby complementing the driver's heightened but potentially overwhelming sensitivity to approaching threats.
% \end{itemize}
\section{Conclusion and Future Work}

We introduced a cognition-to-control framework that models early-stage takeover as boundedly rational decision making and adapts latent cognitive parameters online to forecast short-horizon control quality. By coupling perceptual uncertainty, looming-averse appraisal, and action delay directly to executable steering and pedal commands, the model offers interpretable links between cognitive state and concrete safety outcomes in the first seconds after handover. The online inference layer personalizes these links in real time, which enables assistance that can be both earlier and better targeted.

Empirically, the cognition-adaptive model anticipated hazardous takeovers with higher coverage and longer lead times than baselines. Across 200 collision episodes, it delivered effective warnings at $\geq 0.5\ \mathrm{s}$, $\geq 1\ \mathrm{s}$, and $\geq 2\ \mathrm{s}$ in $89.5\%$, $80.5\%$, and $58.5\%$ of cases, respectively, while a constant-velocity heuristic and a non-adaptive PPO lagged substantially. Beyond behavioral prediction, we assessed the physiological plausibility of the inferred cognitive parameters using real-time eye-tracking data. The estimated perceptual noise parameters exhibited strong temporal alignment with gaze entropy and fixation instability, while the looming-aversion coefficient consistently tracked saccadic bursts and rapid pupil dilation. These findings provide supporting evidence that the inferred parameters capture time-varying shifts in risk perception that are reflected in eye-movement physiology, offering a grounded basis for intervention.

Our findings suggest design opportunities for cognition-aware assistance at handover. When inferred perceptual reliability degrades, systems should slow the interaction, simplify displays to task-critical cues, and enlarge recommended margins. When action delay grows, systems should prioritize buffering interventions such as brief automation hold, brake preload, and feed-forward guidance to bridge delayed human input. Because the model ties each assistance choice to an estimated cause, it can reduce unnecessary overrides while addressing the specific limitation that makes a takeover unsafe.

This work has limitations. The evaluation emphasized freeway work-zone avoidance and simulator-based rollouts, which may not span all traffic geometries or driver strategies. The particle-filter design introduces windowing and process-noise choices that merit sensitivity analysis. While we compared inferred parameters with physiological signals, the current miss rates suggest that some physiological stress responses may be driven by mechanisms not yet captured by the model. Broader evaluation with diverse drivers, vehicles, and sensing stacks remains necessary.

Future work will expand scenario diversity and test closed-loop assistance policies driven by the estimated states to measure user experience, trust, and long-term adaptation. We also plan to investigate the specific influence of different HMI prompts on pre-handover cognition, as our analysis suggested they may induce distinct perceptual shifts. Extending the horizon with risk-sensitive planning, benchmarking against alternative cognitive-computational formalisms, and pursuing track- or on-road studies are important next steps. We also plan to explore fairness diagnostics across driver subgroups to ensure that cognition-aware warnings generalize without unintended performance gaps.

\section{Code Availability}
The code is open-source and available at \url{https://github.com/Jiangxy02/adaptive_rationality}.

\begin{acks}
This research is jointly sponsored by National Natural Science Foundation of China (52125208, 52232015).
\end{acks}

\bibliographystyle{ACM-Reference-Format}
\bibliography{ref}

\end{document}